\documentclass[a4paper,twocolumns,11pt, accepted = 2024-07-22]{quantumarticle}
\pdfoutput=1
\usepackage{graphicx}
\usepackage{multirow}
\usepackage{amsmath}
\usepackage{mathtools}
\usepackage{amssymb}
\usepackage{amsthm}
\usepackage{soul}
\usepackage{verbatim}
\usepackage{bm}
\usepackage[colorlinks=true, linkcolor=blue, citecolor=gray]{hyperref}
\usepackage[table]{xcolor}
\usepackage{tikz}
\usepackage{xspace}
\usepackage{import}
\usepackage[caption=false]{subfig}
\usepackage{bbold}
\usepackage{enumitem}
\usepackage{physics}
\usepackage[numbers,sort&compress]{natbib}

\makeatletter
\newcommand\footnoteref[1]{\protected@xdef\@thefnmark{\ref{#1}}\@footnotemark}
\makeatother

\newcommand{\bs}{\boldsymbol}

\newtheorem{theorem}{Theorem}
\newtheorem{definition}[theorem]{Definition}
\newtheorem{lemma}[theorem]{Lemma}

\newtheorem{algorithm}[theorem]{Algorithm}

\newcommand{\thm}[1]{\hyperref[thm:#1]{Theorem~\ref*{thm:#1}}}
\newcommand{\defn}[1]{\hyperref[defn:#1]{Definition~\ref*{defn:#1}}}
\newcommand{\lem}[1]{\hyperref[lem:#1]{Lemma~\ref*{lem:#1}}}
\newcommand{\prop}[1]{\hyperref[prop:#1]{Proposition~\ref*{prop:#1}}}
\newcommand{\fig}[1]{\hyperref[fig:#1]{Figure~\ref*{fig:#1}}}
\newcommand{\tab}[1]{\hyperref[tab:#1]{Table~\ref*{tab:#1}}}
\renewcommand{\sec}[1]{\hyperref[sec:#1]{Appendix~\ref*{sec:#1}}}
\newcommand{\append}[1]{\hyperref[append:#1]{Appendix~\ref*{append:#1}}}
\newcommand{\cor}[1]{\hyperref[cor:#1]{Corollary~\ref*{cor:#1}}}
\newcommand{\obs}[1]{\hyperref[obs:#1]{Observation~\ref*{obs:#1}}}
\newcommand{\alg}[1]{\hyperref[alg:#1]{Algorithm~\ref*{alg:#1}}}

\newcolumntype{x}[1]{>{\centering\arraybackslash\hspace{0pt}}p{#1}}

\begin{document}
\title{The advantage of quantum control in many-body Hamiltonian learning}
\author{Alicja Dutkiewicz}
\email{dutkiewicz@lorentz.leidenuniv.nl}
\affiliation{Google Quantum AI, Munich 80636, Germany}
\affiliation{Instituut-Lorentz, Universiteit Leiden, 2300RA Leiden, The Netherlands}

\author{Thomas E. O'Brien}
\email{teobrien@google.com}
\affiliation{Google Quantum AI, Munich 80636, Germany}
\affiliation{Instituut-Lorentz, Universiteit Leiden, 2300RA Leiden, The Netherlands}

\author{Thomas Schuster}
\email{tsschuster@berkeley.edu}
\affiliation{Google Quantum AI, Munich 80636, Germany}
\affiliation{Department of Physics, University of California, Berkeley, California 94720 USA}


\begin{abstract}
We study the problem of learning the Hamiltonian of a many-body quantum system from experimental data.
We show that the rate of learning depends on the amount of control available during the experiment.
We consider three control models: one where time evolution can be augmented with instantaneous quantum operations, one where the Hamiltonian itself can be augmented by adding constant terms, and one where the experimentalist has no control over the system's time evolution.
With continuous quantum control, we provide an adaptive algorithm for learning a many-body Hamiltonian at the Heisenberg limit: $T = \mathcal{O}(\epsilon^{-1})$, where $T$ is the total amount of time evolution across all experiments and $\epsilon$ is the target precision. This requires only preparation of product states, time-evolution, and measurement in a product basis.
In the absence of quantum control, we prove that learning is standard quantum limited, $T = \Omega(\epsilon^{-2})$, for large classes of many-body Hamiltonians, including any Hamiltonian that thermalizes via the eigenstate thermalization hypothesis.
These results establish a quadratic advantage in experimental runtime for learning with quantum control.
\end{abstract}

\maketitle

The characterization of unknown systems is a critical topic in science and engineering.
Quantum mechanical systems are governed by a Hamiltonian that determines the evolution of the system in time.
In this setting, system characterization takes the form of learning parameters of the Hamiltonian from experimental data~\cite{baumgratz2016quantum, dutt2021active,ferrie2013best,sergeevich2011characterization,pang2014quantum,yuan2015optimal,fraisse2017enhancing,hou2019control, young2009optimal,sbahi2022provably,wang2017experimental,wiebe2014hamiltonian,wiebe2015quantum, valenti2019hamiltonian, hangleiter2021precise,silva2011practical,boixo2008parameter,shabani2011estimation,zhang2014quantum,wang2017quantum,krastanov2019stochastic,evans2019scalable,li2020hamiltonian,kokail2021quantum,zubida2021optimal, rattacaso2023high, kura2018finite, huang2022learning,yu2022practical,anshu2021sample, haah2021optimal,wang2015hamiltonian, gu2022practical,difranco2009hamiltonian,burgarth2011indirect,bairey2019learning, garrison2018does, qi2019determining, liu2017quantum, kiukas2017remote, stenberg2014efficient, obrien2021quantum,holzapfel2015scalable}.
Hamiltonian learning is of central interest for both applications of quantum technologies and explorations of quantum systems in nature; these include quantum metrology~\cite{sergeevich2011characterization,ferrie2013best,  dutt2021active, baumgratz2016quantum}, molecular structure identification~\cite{sels2020quantum,obrien2021quantum,seetharam2021digital}  quantum device benchmarking~\cite{wiebe2014hamiltonian,wiebe2015quantum, valenti2019hamiltonian}, and  verification of quantum simulations~\cite{carrasco2021theoretical, silva2011practical, hangleiter2021precise}.
A universal objective is to develop learning strategies that consume as few resources (for example, as little time) as possible.
This optimization requires designing both a set of experiments and an algorithm to infer the Hamiltonian from the experimental data.

A central goal of Hamiltonian learning is the so-called Heisenberg limit.
Arising from the quantum Fisher information~\cite{wootters1981statistical,braunstein1994statistical,braunstein1996generalized}, the Heisenberg limit stipulates that the error, $\epsilon$, of any estimation of a Hamiltonian parameter scales at best with the inverse of the total experimental runtime, $T$~\cite{giovanetti2006quantum}.
Protocols that achieve the Heisenberg limit, such as Ramsey spectroscopy~\cite{ramsey1950molecular}, gate-set tomography~\cite{nielsen2021gate}, and Floquet calibration~\cite{google2020Observation}, have become standard procedures across quantum technologies.
Theoretical progress has laid rigorous foundations for Heisenberg-limited learning in single- and few-qubit systems, including theoretical bounds on the learnability and optimal measurement schemes in noiseless~\cite{sergeevich2011characterization,ferrie2013best,pang2014quantum,yuan2015optimal,fraisse2017enhancing, hou2019control, dutt2021active, baumgratz2016quantum, sbahi2022provably, young2009optimal} and noisy~\cite{wan2022bounds, zhou2018achieving} situations.
These works tie in closely with work on Heisenberg-limited phase estimation~\cite{higgins2009demonstrating,kimmel2015robust,dutkiewicz2022heisenberg,lin2022heisenberg}, and Heisenberg-limited unitary estimation~\cite{haah2023query}.

Despite this progress, achieving Heisenberg-limited learning of \emph{many-body} Hamiltonians remains a largely open direction.
The past decade has seen an explosion of learning protocols for many-body Hamiltonians that utilize full~\cite{boixo2008parameter, young2009optimal, shabani2011estimation,zhang2014quantum,wang2017quantum,krastanov2019stochastic,evans2019scalable,li2020hamiltonian,kokail2021quantum,zubida2021optimal, kura2018finite, sbahi2022provably, rattacaso2023high, dutt2021active, huang2022learning} or restricted~\cite{difranco2009hamiltonian,burgarth2011indirect,bairey2019learning, gu2022practical, baumgratz2016quantum, valenti2019hamiltonian, hangleiter2021precise} tomographic access, access to eigenstates~\cite{garrison2018does, qi2019determining, bairey2019learning, evans2019scalable} or thermal states~\cite{haah2021optimal, anshu2021sample, kokail2021quantum, sbahi2022provably, gu2022practical, yu2022practical} of the unknown Hamiltonian, or the ability to copy experimental quantum data to a trusted quantum register~\cite{wiebe2014hamiltonian,wiebe2015quantum,wang2017experimental}.
In most of these works, the estimation error scales as the inverse square root of the total evolution time, known as the standard quantum limit or shot noise limit, and the Heisenberg limit has neither been achieved nor proven impossible.
Interestingly, the only work to achieve the Heisenberg limit thus far, Ref.~\cite{huang2022learning}, requires a high degree of \emph{control} over the system to be learned.
Specifically, the algorithm in Ref.~\cite{huang2022learning}  resembles dynamical decoupling, and requires interleaving ever-smaller steps of time evolution with large single-qubit rotations.
(See also earlier related algorithms~\cite{boixo2008parameter,wang2015hamiltonian}.)
By contrast, the standard quantum limit can be achieved with only the ability to prepare product states, perform time evolution, and measure in a product basis~\cite{haah2021optimal}.
It is natural, and practical, to wonder whether a large degree of control is necessary to achieve Heisenberg-limited learning in the many-body setting.
In this work, we establish a rigorous separation in many-body Hamiltonian learning between systems with and without quantum control.
We consider two forms of quantum control: continuous control, in which one can continuously time-evolve under a Hamiltonian that is the sum of both unknown terms and controlled known terms, and discrete control, in which one can interleave time-evolution under the unknown Hamiltonian with discrete quantum gates.
We begin by providing a new algorithm for learning many-body Hamiltonians, which uses continuous quantum control to achieve Heisenberg-limited learning of any bounded-degree many-body Hamiltonian.
Our algorithm calls an algorithm by Haah, Kothari and Tang (henceforth the ``HKT algorithm'')~\cite{haah2021optimal} as a subroutine, and augments this with quantum controls that simulate reversed time-evolution under a best estimate of the unknown Hamiltonian.
The estimate is updated adaptively as the algorithm proceeds, leveraging techniques for phase estimation introduced in Ref.~\cite{kimmel2015robust}.
Complementary to this algorithm, we show that learning at the Heisenberg limit is \emph{not} possible in the absence of quantum control for large classes of many-body Hamiltonians.
We first show that, if there exist infinitesimal translations in parameter space that do not change the spectrum of the Hamiltonian, then the Hamiltonian parameters cannot be learned at the Heisenberg limit without quantum control.
This result applies to several common classes of Hamiltonians, including qubits or fermions with local or non-local interactions, and even a single qubit evolving under a field along an unknown axis.
We then provide an analogous no-go result for any many-body Hamiltonian that satisfies the eigenstate thermalization hypothesis (ETH), a widely validated conjecture for how closed many-body systems relax to effectively thermal states~\cite{srednicki1994chaos,rigol2008thermalization,d2016quantum,deutsch2018eigenstate}.
Both of our no-go theorems apply quite generally, and preclude any algorithm without sufficient quantum control, including adaptive algorithms, algorithms with access to a quantum memory, and algorithms that involve arbitrarily complex quantum or classical operations that do not involve the unknown Hamiltonian.
This generality arises naturally because our proofs utilize bounds on the quantum Fisher information~\cite{wootters1981statistical,braunstein1994statistical,braunstein1996generalized} and related quantities that we define for unitary quantum processes.

\section{Background and problem definition}\label{sec:background}
\begin{figure*}
    \centering
    \includegraphics[width=0.6\linewidth]{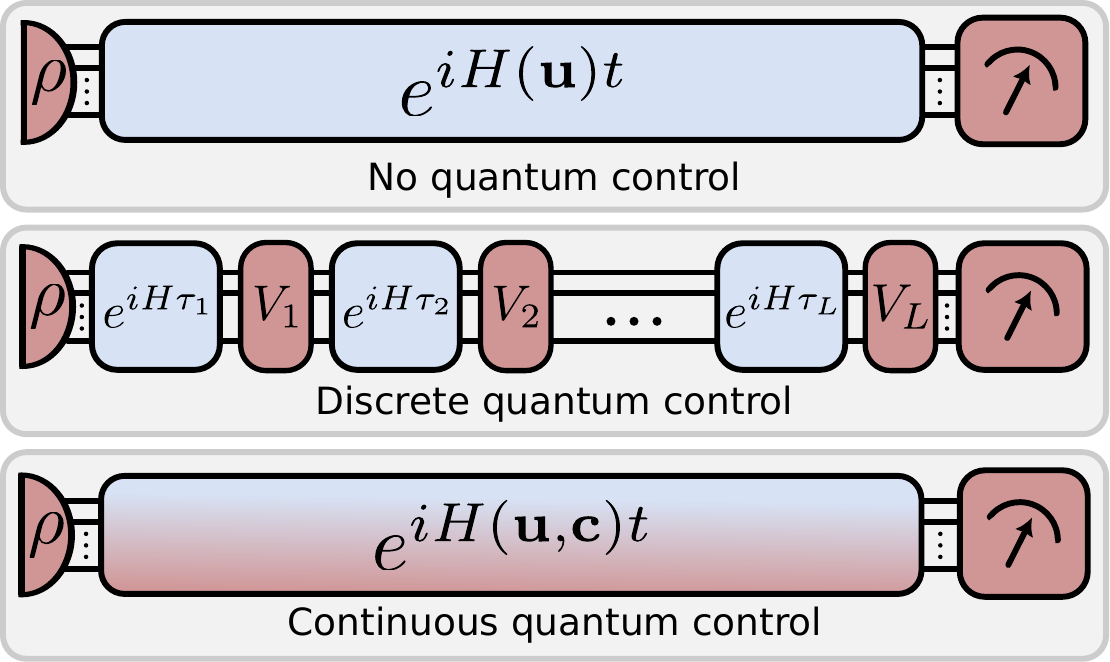}
    \caption{Schematic quantum circuit diagrams of the three experimental models that we consider in this work.
    Colors denote when the experimentalist has complete control (red), some control (light blue-red gradient), or no control (light blue) over the operation.
    We assume the experimentalist has the freedom to specify state preparation (half-ellipse) and measurement (dial symbol).
    Further, during time evolution (rectangles) we assume the freedom to specify the evolution times $t / \tau_l$, and the instantaneous quantum operations $V_l$ (in the discrete quantum control model), or the control parameters $\mathbf{c}$ (in the continuous quantum control model).
    In the discrete control model, we assume the quantum operations $V_l$ are unitary but allow them to act on an arbitrary number of ancilla qubits (not shown), and we hide the dependence of $H$ on the unknown parameters $\mathbf{u}$ for clarity.}
    \label{fig:learning_schematic}
\end{figure*}

We consider the problem of learning a Hamiltonian,
\begin{equation} \label{eq:Ham_no_control}
    H(\mathbf{u}) = \sum_{a=1}^{N_p} u_a P_a, \hspace{1cm} |u_a|\leq 1
\end{equation}
where $u_a$ are unknown parameters and $P_a$ are $k$-local Pauli operators on $N$ qubits, with $k = \mathcal{O}(1)$.
We abbreviate $\mathbf{u} = (u_1,\ldots,u_{N_p})$.
Our goal is to learn the parameters $\mathbf{u}$ from a system with dynamics governed by $H(\mathbf{u})$ by performing experiments on the aforementioned system.
Each experiment consists of state preparation, time-evolution, and measurement.
We assume that the time-evolution features some form of `black-box access' to $H(\mathbf{u})$ that we describe below.

We consider three experimental settings  (see Fig.~\ref{fig:learning_schematic}), which are distinguished by the presence of `quantum control': the ability to manipulate the quantum state of the unknown system during time evolution.
(These settings do not change our abilities regarding system preparation or measurement. In each setting, we consider either product state preparation and computational basis measurement -- for our learning algorithm, Theorem~\ref{thm:HHKT_performance_informal}, or arbitrary state preparation and measurement -- for our lower bounds, Theorems~\ref{thm:no_unitary_equivalence_informal},~\ref{thm:eth_learning_informal}.)
First, we consider learning with `no quantum control' (top, Fig.~\ref{fig:learning_schematic}), in which all experiments consist only of state preparation, a single instance of time evolution under the unknown Hamiltonian, $U = e^{-i H(\mathbf{u}) t}$, and measurement.
Unless otherwise stated, we assume the experimentalist can perform arbitrary state preparations and measurements, and evolve for an arbitrary time $t$ (at a cost given in the next paragraph).
Second, we consider experiments with `discrete quantum control' (middle, Fig.~\ref{fig:learning_schematic}), in which time evolution may be interleaved with instantaneous quantum operations by the experimentalist (i.e. operations during which $H(\mathbf{u})$ does not act).
This definition is appropriate for hybrid analog-digital quantum platforms, and for applications of Hamiltonian learning to verify digital quantum simulations~\cite{carrasco2021theoretical, silva2011practical, hangleiter2021precise}.
In practice, the degree of discrete control may be limited (for example, by energetic considerations, pulse discretization, or noise), so we will set an upper bound, $L$, on the number of instantaneous operations in a single experiment~\footnote{Let us elaborate on this point briefly. In many settings of interest, the Hamiltonian to be learned will be native to the system of interest, and it may not be possible to instantaneously halt the Hamiltonian during the implementation of a discrete quantum gate. Thus, the implementation of a digital gate might incur some error (e.g.~if one attempts to implement the gate when the native Hamiltonian is still ``on'') or require some additional time (e.g.~to transition the system between Hamiltonian and digital modes of operation). To capture these effects within a simple theoretical model, we assume that the total number of digital gates is bounded by some number $L$, and study the dependence of learning on $L$.}.
Finally, we consider experiments with `continuous quantum control' (bottom, Fig.~\ref{fig:learning_schematic}), in which control terms are added to the Hamiltonian itself:
\begin{equation} \label{eq:Ham_control}
    H(\mathbf{u},\mathbf{c}) = \sum_{a=1}^{N_p} (u_a + c_a) P_a, \hspace{1cm} |u_a|,|c_a|\leq 1.
\end{equation}
Compared to discrete quantum control, this model reflects the fact that in physical quantum systems the experimental control frequently has a bounded strength that renders instantaneous operations impossible.
For simplicity, we assume the control parameters $c_a$ are time-independent within a single experiment, but can be reset by the experimentalist between different experiments.
In practice, experiments may also be able to modify $c_a$ within the timescale of a single experiment. 
However, this limited degree of control will already be sufficient for our learning algorithm in Theorem~\ref{thm:HHKT_performance_informal}.

To quantify the performance of a Hamiltonian learning algorithm, we now define a cost model and a metric for success.
The cost of generating data is taken to be the total experiment time 
\begin{equation}\label{eq:total_expt_time}
    T=\sum_x t_x,
\end{equation}
where $x$ labels an individual experiment, and $t_x$ is the duration of time-evolution under the unknown Hamiltonian in the experiment.
This neglects any cost of state preparation and measurement.
We quantify the accuracy of an estimator $\tilde{\mathbf{u}}$ of the unknown parameters $\mathbf{u}$ by requiring that the maximum root-mean-square (RMS) error be bounded as
\begin{equation}\label{eq:max_RMS_error}
    \epsilon\leq \max_a\Big[\overline{(\tilde{u}_a-u_a)^2}\Big]^{1/2},
\end{equation}
where the overline denotes an expectation value over experimental outcomes.

The Heisenberg limit provides a universal bound, $T = \Omega(\epsilon^{-1})$, on the total experiment time $T$ required to achieve a maximum RMS error $\epsilon$~\cite{giovanetti2006quantum,higgins2009demonstrating,ferrie2013best}.
(We provide a derivation of this via the quantum Cramer-Rao bound in Appendix~\ref{app:fisher_information}.)
The bound passes through to other error metrics up to factors of $\sqrt{N_p}$ using a median-of-means approach  (see Appendix~\ref{app:alternativeerror}).
The Heisenberg limit applies to any learning model, and does not prohibit stricter bounds being established in specific settings.

\section{Results}\label{sec:results}

In this work, we establish a gap between the learnability of Hamiltonians with and without quantum control.
We present our results informally in this section, and establish rigorous statements and proofs in the Appendix.

Our first result is an algorithm to learn any local many-body Hamiltonian at the Heisenberg limit using continuous quantum control (Appendix~\ref{app:HKT_Heisenberg}).
Our algorithm builds off the the HKT algorithm introduced in Ref.~\cite{haah2021optimal}, which learns a many-body Hamiltonian at the \emph{standard quantum limit}, $T = \Omega(\epsilon^{-2})$, from experiments involving no quantum control.
The HKT algorithm is standard quantum limited because each experiment involves evolution only up to a maximum time $t_c$, which is constant in $\epsilon$ (see Appendix~\ref{app:HKT} for details).
This restriction is fundamental to the HKT algorithm, which involves approximating time-evolved operators by their Taylor series, $e^{iHt} P_a e^{-iHt} \approx P_a + i t [H, P_a] + \mathcal{O}(t^2)$, which diverges at large $t$.

\begin{figure*}
    \centering
    \includegraphics[width=0.7\linewidth]{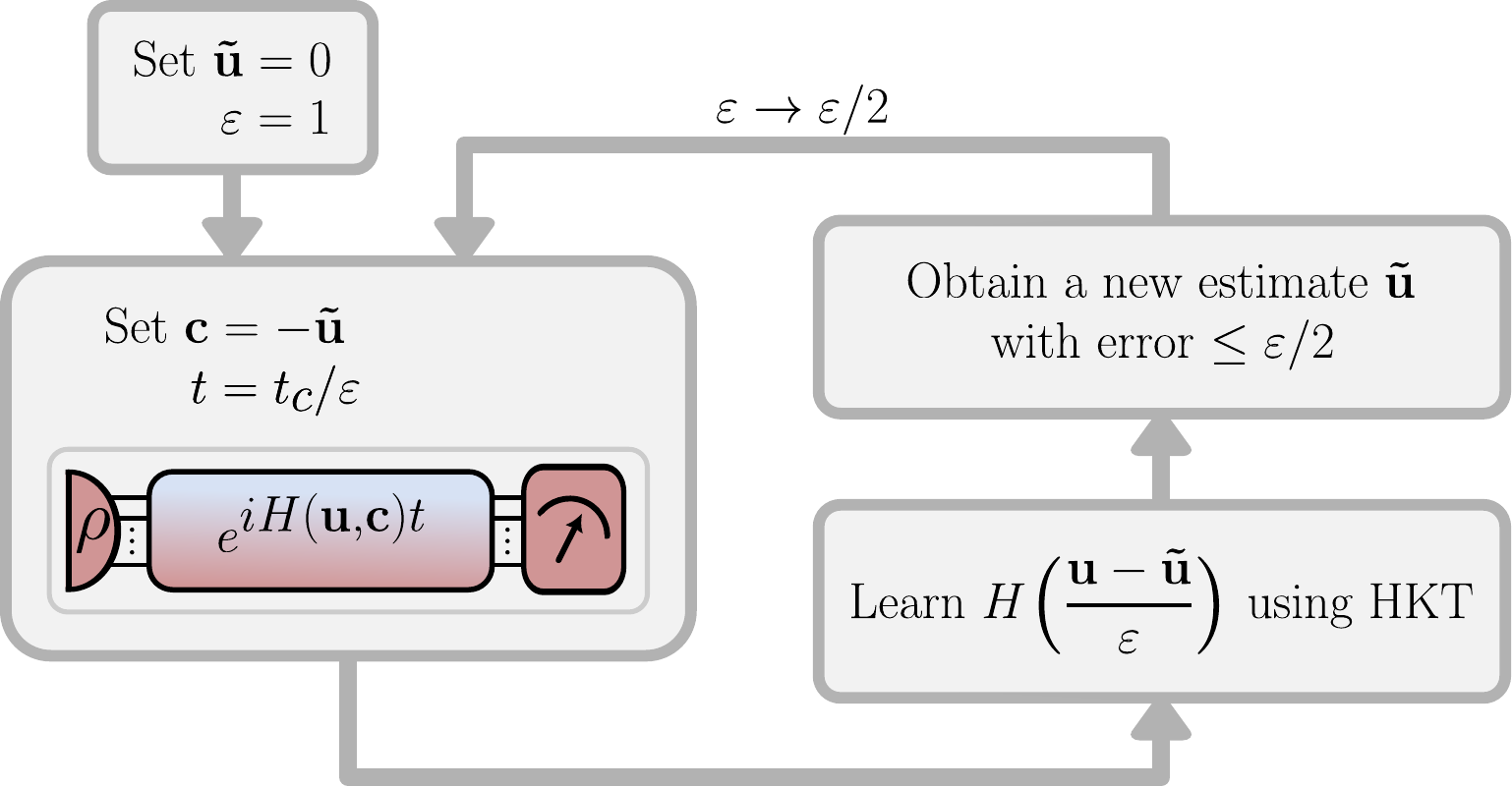}
    \caption{Flowchart of our Heisenberg-limited algorithm. The algorithm starts with an initial guess $\mathbf{\tilde u}$ for the unknown Hamiltonian parameters $\mathbf{u}$ accurate up to $\varepsilon =1$, and adaptively adjusts the control parameters to refine the estimate. The HKT algorithm of Ref. \cite{haah2021optimal} is used to learn a rescaled Hamiltonian (Eq.~\eqref{eq:rescaled_hamiltonian}) and then the estimate of $\mathbf{u}$ is updated based on the learned Hamiltonian. The algorithm continues iteratively until the error $\varepsilon$ in the estimate is smaller than a specified threshold~$\epsilon$.}
    \label{fig:HHKT_flowchart}
\end{figure*}

We surpass the standard quantum limit by introducing adaptive quantum controls to the HKT algorithm, as illustrated by the flowchart in Fig.~\ref{fig:HHKT_flowchart}. 
Suppose that previous experiments allow us to form an estimate $\tilde{\mathbf{u}}$ of $\mathbf{u}$ with error less than $\varepsilon$.
We then set our continuous quantum control parameters to the negative of the estimated Hamiltonian, $\mathbf{c}=-\tilde{\mathbf{u}}$.
This yields time-evolution under the Hamiltonian,
\begin{align}
    H(\mathbf{u},\mathbf{c}) &= \varepsilon H(\mathbf{u}'),\hspace{0.5cm} H(\mathbf{u}') = \sum_a u'_a P_a,\\
    u'_a &= (u_a-\tilde{u}_a)/\varepsilon,\label{eq:rescaled_hamiltonian}
\end{align}
where the new parameters obey $|u'_a|\leq 1$.
We can now apply the HKT algorithm to the rescaled Hamiltonian $H'$, by evolving under the original Hamiltonian $H$ for a time $t_c/\varepsilon$.
This improves the error in our estimate of $\mathbf{u}$ by a constant factor.
Repeating this procedure in a similar fashion to protocols for Heisenberg-limited phase estimation~\cite{kimmel2015robust,dutkiewicz2022heisenberg} yields our result:
\begin{theorem}[informal]\label{thm:HHKT_performance_informal}
    Consider a low-intersection Hamiltonian $H(\mathbf{u})$ with unknown parameters $\mathbf{u}$ and continuous quantum control as in Eq.~(\ref{eq:Ham_control}). There exists an algorithm to learn $\mathbf{u}$ within maximum RMS error $\epsilon$ using a total experimental time $T= \mathcal{O}(\epsilon^{-1})$ and only product states and computational basis measurements.
\end{theorem}
\noindent (The term `low-intersection' is a technical constraint inherited from the HKT algorithm that we describe in Appendix~\ref{app:HKT}.) Our algorithm bears some resemblance to that of Ref.~\cite{kura2018finite}; however, by using the HKT algorithm as an inner loop, we improve the error bound exponentially in $N$ and  remove the need for entanglement with a large ancilla register. 

Our remaining results place lower bounds, $T = \Omega(\epsilon^{-2})$, on Hamiltonian learning without control.
Our first bound is particularly simple: suppose two Hamiltonians $H(\mathbf{u})$ and $H(\mathbf{u} + \delta\mathbf{u})$ are equal up to a unitary transformation. Then, they cannot be distinguished at the Heisenberg limit for $\epsilon\approx\|\delta\mathbf{u}\|_{2}$ without quantum control (Appendix~\ref{app:no_control_no_heisenberg}).
This is based on a straightforward calculation from the quantum Fisher information which shows that the eigenvalues of $H(\mathbf{u})$ can be learned at the Heisenberg limit, but the eigenvectors cannot.
Our Hamiltonian learning problem does not target the learning of eigenvalues or eigenvectors, but rather the unknown parameters $\mathbf{u}$.
However, we find that if there exists a shift in the unknown parameters that leaves the eigenvalues unchanged, then the parameters cannot be learned at the Heisenberg limit:
\begin{theorem}[informal]\label{thm:no_unitary_equivalence_informal}
    Consider a Hamiltonian $H(\mathbf{u})$ with unknown parameters $\mathbf{u}$, and either no quantum control or discrete quantum control with a constant number $L = \mathcal{O}(1)$ of interleaved operations.
    Suppose that there exists an infinitesimal transformation, $\mathbf{u} \rightarrow \mathbf{u} + \delta \mathbf{u}$, that leaves the spectrum of the Hamiltonian unchanged to leading order in $\delta \mathbf{u}$. 
    Then learning the linear combination of parameters $\mathbf{u} \cdot (\delta \mathbf{u} / \lVert \delta \mathbf{u} \rVert)$ to within RMS error $\epsilon$ requires a total experimental time $T = \Omega(\epsilon^{-2})$
\end{theorem}
\noindent 
Theorem~\ref{thm:no_unitary_equivalence_informal} is completely general, in that it holds for arbitrary state preparation and measurement in the experiment.
As stated, the theorem concerns the RMS error of a linear combination of parameters specified by $\delta \mathbf{u}$. 
If $\delta \mathbf{u}$ has support on only a single parameter, then this trivially lower bounds our original choice of error metric, the maximum RMS error over the individual parameters $u_a$ [Eq.~(\ref{eq:max_RMS_error})].
More generally, the error in any linear combination lower bounds the maximum RMS error up to a factor of $\sqrt{1/N_p}$. 

Hamiltonians that obey the conditions of Theorem~\ref{thm:no_unitary_equivalence_informal} are quite common in the natural world.
For instance the single-qubit Hamiltonian,
\begin{equation}
    H(\mathbf{u})=u_xX + u_yY + u_zZ,~\label{eq:single_qubit_H}
\end{equation}
has the same spectrum whenever $\|\mathbf{u}_1\|_2=\|\mathbf{u}_2\|_2$.
The angle of $\mathbf{u}$ can therefore not be learned at the Heisenberg limit.
This point was established by other methods in Refs.~\cite{hou2019control, baumgratz2016quantum}; Theorem~\ref{thm:no_unitary_equivalence_informal} can be thus seen as a generalization of the single-qubit no-go theorems in these works to the general many-body case.
Similarly, the fermionic Hamiltonian,
\begin{equation}
    H(\mathbf{u})=\sum_{ij}u^{(1)}_{ij}c^{\dag}_ic_j + \sum_{ijkl}u^{(2)}_{ijkl}c^{\dag}_ic^{\dag}_jc_kc_l,
\end{equation}
possesses orbits of $\mathbf{u}$ connected by Givens rotations~\cite{wecker2015solving}.
This holds for both local and all-to-all connectivity, since Givens rotations may be chosen to preserve locality.
Analogous spin Hamiltonians are discussed in Appendix~\ref{app:no_control_no_heisenberg}.

While the applicability of Theorem~\ref{thm:no_unitary_equivalence_informal} is broad, the conditions of the theorem can be circumvented by adding only a small amount of continuous quantum control or prior knowledge about the Hamiltonian (for example, by adding fixed non-adaptive background fields~\cite{yuan2015optimal,fraisse2017enhancing,hou2019control,google2020Observation}).
Our second lower bound excludes Heisenberg-limited learning on a more robust physical context.
Namely, we show that learning is bounded by the standard quantum limit, $T = \Omega(\epsilon^{-2})$, for any Hamiltonian that thermalizes via the eigenstate thermalization hypothesis (ETH)~\cite{srednicki1994chaos,rigol2008thermalization,d2016quantum,deutsch2018eigenstate} (Appendix~\ref{app:ETH_no_learning}).
Thermalization describes the relaxation in time of local observables to their thermal values~\cite{d2016quantum}.
The connection between learning without quantum control and thermalization is natural, since both concern the late time dynamics of the unitary evolution $e^{-iH(\mathbf{u})t}$.
Thermalization clearly precludes learning at the Heisenberg limit from local measurements, since they approach thermal distributions that are constant in time.
However, it is not a priori clear whether this extends to more complex measurement strategies, such as those involving backwards time-evolution under an estimate of the Hamiltonian~\cite{wiebe2014hamiltonian,wiebe2015quantum}.
Indeed, even thermalizing Hamiltonians exhibit coherent late time phenomena in their own eigenbasis, which suggests that some degree of Heisenberg limited learning may be possible.

We resolve these questions by invoking the ETH, a conjecture for how thermalization occurs in closed Hamiltonian systems.
The ETH concerns the matrix elements of local operators in the Hamiltonian eigenbasis.
It posits that (i) the diagonal matrix elements are equal to the expectation value of the operator in a corresponding thermal state, and (ii) the off-diagonal matrix elements have magnitude exponentially small in the system size, and are random up to certain macroscopic constraints~\cite{srednicki1994chaos,rigol2008thermalization,d2016quantum,deutsch2018eigenstate}. 
We formulate these conditions precisely in  Appendix~\ref{app:ETH}.
The core tenets of the ETH have been thoroughly validated by extensive numerical studies across many Hamiltonians; see Ref.~\cite{d2016quantum} for a comprehensive review.
Taking the validity of ETH as an assumption, we establish the following theorem:
\begin{theorem}[informal]\label{thm:eth_learning_informal}
Consider a Hamiltonian $H(\mathbf{u})$ with an extensive number, $N_p = \Omega(N)$, of unknown parameters $\mathbf{u}$, and either no quantum control or discrete quantum control with a constant number, $L = \mathcal{O}(1)$, of interleaved operations.
Suppose that, within the states accessed by an experiment, the Hamiltonian obeys the ETH and that its thermal expectation values have bounded derivatives with respect to the temperature.
Then learning the parameters $\mathbf{u}$  with maximum RMS error $\epsilon$ requires a total experimental time, $T = \Omega(\epsilon^{-2})$, as long as the time is sub-exponential in the system size, $T = o(\text{exp}(N))$.
\end{theorem}
\noindent The theorem holds for arbitrary state preparation and measurement, so long as the Hamiltonian is thermalizing on the states accessed during the experiment.
The required condition on derivatives of thermal expectation values is quite loose, and is a standard feature of the thermodynamic limit.
(In Appendix~\ref{app: bounded derivative}, we prove it holds for geometrically local Hamiltonians whenever correlations decay exponentially~\cite{kliesch2014locality,bluhm2022exponential}.)
The restriction to sub-exponential times is natural, since at later times the system may exhibit non-thermal revivals that are not captured by the ETH.

Our proof of Theorem~\ref{thm:eth_learning_informal} relies on several technical lemmas, which may be of interest for future work on multi-parameter Hamiltonian learning or rigorous analysis of thermalizing Hamiltonians.
We begin by expressing the quantum Fisher information in terms of time-integrals over operators entering the Hamiltonian, which are then amenable to analysis via the ETH.
We establish that the spectral norm of the contribution of off-diagonal matrix elements of these operators is bounded above by $\mathcal{O}(t^{1/2})$, and hence cannot lead to learning at the Heisenberg limit.
Our proof leverages techniques from random matrix theory.
Meanwhile, the diagonal matrix elements grow instead as $\mathcal{O}(t)$. However, we show that their contribution to the Fisher information matrix is low-rank when the derivatives of thermal expectation values are bounded.
Intriguingly, this rank-deficiency is only a limitation for  \emph{many}-parameter learning. In keeping with our initial intuition, we find that Heisenberg-limited learning is possible for many-body Hamiltonians where only a constant number of parameters are unknown, using non-local measurement strategies (Appendix~\ref{app:ETH_single_parameters}).

\section{Comparison to previous work}

\textbf{Models of quantum control:} Prior studies on Hamiltonian learning have considered a diverse range of experimental scenarios, almost all of which are encapsulated by one of our three models of quantum control. 
Many works fall under our `no quantum control' model~\cite{boixo2008parameter, burgarth2011indirect, difranco2009hamiltonian, dutt2021active, ferrie2013best, franca2022efficient, gu2022practical, haah2021optimal, hangleiter2021precise, li2020hamiltonian, obrien2021quantum, rattacaso2023high, sergeevich2011characterization, shabani2011estimation, wang2017experimental, wang2017quantum, wiebe2014hamiltonian, young2009optimal, yu2022practical, zhang2014quantum, zubida2021optimal,holzapfel2015scalable}, some of which include additional, experimentally-motivated constraints that are not imposed in our work~\cite{burgarth2011indirect, shabani2011estimation, li2020hamiltonian, hangleiter2021precise}.
Notably, our no control model encompasses previous proposals for ``quantum-assisted'' Hamiltonian learning~\cite{wiebe2014hamiltonian,wiebe2015quantum,wang2017experimental, kokail2021quantum,holzapfel2015scalable}, which nonetheless use only a single application of time-evolution under the unknown Hamiltonian.
A comparatively fewer number of works fall under our discrete quantum control model~\cite{kura2018finite, huang2022learning, yuan2015optimal, hou2019control, liu2017quantum}; these are referred to as ``quantum-enhanced'' experiments in Ref.~\cite{huang2022learning}.
Meanwhile, our continuous quantum control model encapsulates a number of additional works that use both time-independent \cite{fraisse2017enhancing, kura2018finite, stenberg2014efficient, valenti2019hamiltonian} and time-dependent continuous control~\cite{bairey2019learning, liu2017quantum, kiukas2017remote, liu2017control, krastanov2019stochastic, pang2017optimal}.
Some works have allowed evolving an entangled state over multiple copies of the system~\cite{baumgratz2016quantum, kura2018finite}, which are encapsulated within our discrete control model by allowing swap operations between the the system and an ancillary register~\cite{sekatski2017quantum}.
Finally, a number of other works have focused on learning from access to steady states of Hamiltonians (e.g. eigenstates or thermal states)~\cite{haah2021optimal, anshu2021sample, sbahi2022provably, gu2022practical,garrison2018does, qi2019determining,bairey2019learning, evans2019scalable}; while in principle steady states might be obtained by using time-evolution under the unknown Hamiltonian, this is largely orthogonal to the focus of our work.

\textbf{Few-parameter Hamiltonian learning:} Many works have investigated learning Hamiltonians with a single~\cite{boixo2008parameter, sergeevich2011characterization,fraisse2017enhancing, hou2019control, liu2017quantum, yuan2015optimal} or few~\cite{baumgratz2016quantum, ferrie2013best, stenberg2014efficient, young2009optimal,krastanov2019stochastic, kura2018finite, wang2017quantum,bairey2019learning, bairey2019learning, difranco2009hamiltonian, dutt2021active, hangleiter2021precise, kiukas2017remote, obrien2021quantum, qi2019determining, valenti2019hamiltonian, wang2015hamiltonian, wang2017experimental, liu2017control} unknown parameters. Here, we also use ``few'' to refer to algorithms whose scaling with the number of unknown parameters is either exponential~\cite{krastanov2019stochastic, kura2018finite, wang2017quantum} or unknown~\cite{bairey2019learning, bairey2019learning, difranco2009hamiltonian, dutt2021active, hangleiter2021precise, kiukas2017remote, obrien2021quantum, qi2019determining, valenti2019hamiltonian, wang2015hamiltonian, wang2017experimental, liu2017control}.
Several of these works have also considered the benefits of quantum control in Hamiltonian learning~\cite{boixo2008parameter, sergeevich2011characterization,fraisse2017enhancing, hou2019control, liu2017quantum, yuan2015optimal,baumgratz2016quantum, ferrie2013best, stenberg2014efficient, young2009optimal}. 
For the simplest such Hamiltonian, with $H(u) = uP$ and $u$ unknown, learning corresponds to the well-known phase estimation problem.
In this case, the Heisenberg limit can be reached with no quantum control~\cite{higgins2009demonstrating}, and thus quantum control yields no asymptotic advantage~\cite{liu2017quantum}.
However, beyond this, there are known examples of Hamiltonians with a single unknown parameter where learning without quantum control is standard quantum limited~\cite{yuan2015optimal}.
Previous works have proven that the Heisenberg limit can be achieved in these scenarios by leveraging  discrete~\cite{yuan2015optimal} or continuous~\cite{fraisse2017enhancing} quantum control.
Some of these algorithms leverage approximate backwards time-evolution~\cite{yuan2015optimal,fraisse2017enhancing}, in a similar manner to our many-body learning algorithm.

\textbf{Standard-quantum-limited learning for many-body Hamiltonians:} Rigorous algorithms for learning many-body Hamiltonians in polynomial time have been demonstrated in Refs.~\cite{anshu2021sample, burgarth2011indirect, evans2019scalable, franca2022efficient, gu2022practical, gu2022practical, haah2021optimal, haah2021optimal, li2020hamiltonian, rattacaso2023high, sbahi2022provably, shabani2011estimation, wiebe2014hamiltonian, yu2022practical, zhang2014quantum, zubida2021optimal}, all with precision scaling as the standard quantum limit.
For other algorithms~\cite{dutt2021active} the Heisenberg limit is anticipated, but has not been proven.
Among these, the algorithm of Ref.~\cite{haah2021optimal} is crucial to the present manuscript, as we use it as a subroutine in our Heisenberg-limited learning algorithm.

\textbf{Heisenberg-limited learning for many-body Hamiltonians:} The only prior work to achieve the Heisenberg limit in many-body Hamiltonian learning was Ref.~\cite{huang2022learning}.
Their algorithm interleaves large single-qubit rotations in between short steps of time-evolution, in a manner similar to dynamical decoupling.
In particular, the rate at which single-qubit gates are applied becomes ever more frequent with the desired precision $\epsilon$, such that the total algorithm requires a number of interleaved operations $L = \Theta(\epsilon^{-2})$.
Interestingly, this is quadratically worse than our lower bound, $L=\Omega(\epsilon^{-1})$. It remains an open question whether the algorithm in Ref.~\cite{huang2022learning} can be strengthened to match this lower bound.
In terms of experimental feasibility, for small $\epsilon$ the frequent interleaved operations in Ref.~\cite{huang2022learning} may be difficult to apply in practice (depending, for example, on the power of single-qubit pulses relative to the strength of the unknown Hamiltonian).
In comparison, our algorithm only uses control operations of the same strength as the unknown Hamiltonian.
At the same time, these controls must be able to implement backwards time-evolution under a best estimate of the unknown Hamiltonian, which will require  two-qubit interactions or gates.
Thus, where each algorithm succeeds in practice will depend on the specific control set of the system of interest.

\section{Discussion}\label{sec:discussion}

The above results establish an advantage in many-body Hamiltonian learning for both continuous and discrete quantum control.
For the former, the algorithm in Theorem~\ref{thm:HHKT_performance_informal} establishes that any $k$-local Hamiltonian can be learned at the Heisenberg limit with continuous quantum control.
This circumvents the no-go Theorems~\ref{thm:no_unitary_equivalence_informal} and~\ref{thm:eth_learning_informal}, and establishes an advantage for continuous quantum control over the no quantum control model in learning Hamiltonians where either of these theorems apply.
Moreover, the total \emph{number} of experiments needed to implement our algorithm is only $\mathcal{O}(\log \epsilon^{-1})$.
This is exponentially fewer than all known learning algorithms with no quantum control. 
At the same time, our algorithm requires the ability to individually modify every term in the Hamiltonian, a relatively strong requirement for physical experiments.
A clear question for future research is whether this can be relaxed, for example to experiments with control over only single-qubit terms or global combinations of terms.
This is relevant in the case of e.g. NMR~\cite{sels2020quantum,obrien2021quantum,seetharam2021digital} where one has only control over a global magnetic field.

For discrete quantum control, our no-go theorems combine with the algorithm introduced in Ref.~\cite{huang2022learning} to establish a separation in Hamiltonian learning as a function of the number $L$ of interleaved quantum operations.
Ref.~\cite{huang2022learning} establishes a Heisenberg-limited algorithm in the discrete quantum control setting that implements a dynamical decoupling scheme requiring $L=\Theta(T^2)$ interleaved unitaries.
Meanwhile, our formal versions of Theorems~\ref{thm:no_unitary_equivalence_informal} and~\ref{thm:eth_learning_informal} establish that $L$ must scale at least linearly with the evolution time, $L = \Omega(T)$  (see Appendix~\ref{app:no_control_no_heisenberg} and~\ref{app:ETH_no_learning}).
Given recent work establishing a Heisenberg-limited unitary learning scheme~\cite{haah2023query}, we expect that Heisenberg-limited Hamiltonian learning in the discrete quantum control model with $L=\Theta(T)$ can be achieved, making the bound tight.
This would require repeatedly alternating forward time-evolution by $e^{iHdt}$ and backward time evolution by $e^{-iH(\tilde{u})dt}$ for $dt$ constant in $\epsilon$.
This follows a different principle to Ref.~\cite{huang2022learning} (term cancellation versus dynamical decoupling); these two principles correspond to the limits of optimal control for learning in the single-parameter case~\cite{fraisse2017enhancing}.
We think it an interesting open question whether a dynamical decoupling scheme with $L=\Theta(T)$ can be achieved; we speculate that this may indeed be possible by leveraging rigorous approximations for Floquet evolution with geometrically local Hamiltonians~\cite{abanin2017rigorous}.

While each of our no-go theorems apply to Hamiltonians that are common throughout nature, Theorem~\ref{thm:eth_learning_informal} is in some ways more robust than Theorem~\ref{thm:no_unitary_equivalence_informal}.
As we have remarked, even small modifications to the Hamiltonian, such as the addition of an external field of known value, can break the unitary equivalence necessary for Theorem~\ref{thm:no_unitary_equivalence_informal}.
However, Theorem~\ref{thm:eth_learning_informal} cannot be so easily circumvented.
For example, suppose that one has continuous quantum control, but chooses control parameters $\mathbf{c}$ such that each instance of the Hamiltonian $H(\mathbf{u},\mathbf{c})$ satisfies the ETH.
From the proof of Theorem~\ref{thm:eth_learning_informal}, one can show that $\Omega(N/\mathrm{polylog}(N))$ unique choices of $\mathbf{c}$ will be required to learn at the Heisenberg limit.
Our algorithm in Theorem~\ref{thm:HHKT_performance_informal} circumvents this by choosing $\mathbf{c}$ such that the system does not thermalize until ever-later times in each successive experiment.
We conjecture that similar methods might be more broadly required: that efficiently learning many-body Hamiltonians at the Heisenberg limit requires breaking, or delaying, ergodicity.
Interestingly, in Appendix~\ref{app:ETH_single_parameters} we provide strong arguments that Heisenberg-limit learning \emph{can} be achieved in thermalizing Hamiltonians when only a constant $\mathcal{O}(1)$ number of parameters are unknown, using the quantum-assisted Hamiltonian learning algorithms of Refs.~\cite{wiebe2014hamiltonian,wiebe2015quantum,wang2017experimental}.

Interestingly, we can in fact understand our Theorem~\ref{thm:eth_learning_informal} for thermalizing Hamiltonians 
 in terms of Theorem~\ref{thm:no_unitary_equivalence_informal}.
This comes from an observation within the proof of Theorem~\ref{thm:eth_learning_informal} that there exist extensively many linear combinations $\sum_ah_aP_a$ of local operators whose matrix elements are purely off-diagonal in the eigenbasis of the Hamiltonian.
This implies that the spectrum of the Hamiltonian is invariant (to first order in perturbation theory) under the parameter shift, $H\rightarrow H+\lambda \sum_a h_aP_a$, which falls precisely under the conditions of Theorem~\ref{thm:eth_learning_informal}.
(To see this, recall that from perturbation theory the change in an eigenenergy $E_j$ is given by $\lambda\langle E_j| \sum_a h_a P_a |E_j\rangle+\mathcal{O}(\lambda^2)$, which vanishes to first order when $\sum_ah_aP_a$ is off-diagonal.)
The fact that the spectrum of a Hamiltonian obeying the ETH is independent of so many microscopic parameters is a stark illustration of the averaging effects of thermalization.

In summary, we have established separations in Hamiltonian learning between experiments that can control a system throughout time-evolution, and those that cannot.
Our work begins with a rigorous formulation of three models of quantum control: no control, discrete, and continuous.
These definitions encapsulate nearly all previous algorithms for Hamiltonian learning.
Utilizing continuous quantum control, we introduce a new algorithm for learning any many-body Hamiltonian at the Heisenberg limit, which combines techniques from robust phase estimation~\cite{kimmel2015robust} with the HKT algorithm~\cite{haah2021optimal}.
Our algorithm joins with that of Ref.~\cite{huang2022learning} among the first provable algorithms for learning many-body Hamiltonians at the Heisenberg limit.
Complementary to this, we establish that the Heisenberg limit is impossible to achieve in general without quantum control.
We identify two roadblocks---unitary equivalences between Hamiltonians, and thermalization via the eigenstate thermalization hypothesis---and show that either of these preclude nearly all learning strategies that do not interrupt time-evolution at, at least, a constant rate.
We hope that these results provide a new lens through which to view existing Hamiltonian learning algorithms, and spark further refinements in algorithms to come.

\acknowledgments{The authors are grateful to Hsin-Yuan Huang and Vadim Smelyanskiy for valuable discussions and advice on this work, and Robin Kothari for detailed comments on the manuscript.
A.D. is supported by a Google PhD Fellowship.
 T.S. acknowledges support
from the National Science Foundation (QII-TAQS program and GRFP).}

\bibliographystyle{unsrtnat}
\bibliography{bibliography}

\appendix
\onecolumn
\section{Preliminaries}\label{app:background}

We begin in Appendix~\ref{app:formal setup} by providing a formal definition of a quantum experiment, quantum control, and the many-body Hamiltonian learning problem. 
We discuss the relations between various error bounds on Hamiltonian learning in Appendix~\ref{app:alternativeerror}.
In Appendix~\ref{app:fisher_information}, we establish several basic facts regarding Hamiltonian learning experiments, related to the Fisher information.

Throughout this work, we use $\mathcal{O},\Theta, \Omega,o$ and $\omega$ following standard `big-$O$' notation.
We use $\|\cdot\|_2$ and $\|\cdot\|_{\infty}$ to refer to the standard vector $2$- and $\infty$-norms over $\mathbb{C}$ and $\mathbb{R}$, and $\|\cdot\|_s$ to refer to the spectral norm of a matrix or operator.
We use $\mathbb{P}^N=\{I,X,Y,Z\}^N$ to denote the set of all tensor products of Pauli matrices on $N$ qubits.
We use bold-font to imply that the object in question is a vector of dimension equal to the number of parameters $N_p$ in our learning problem; depending on the context this may be a vector of real or imaginary numbers, or a vector of operators acting on the $N$-qubit Hilbert space $\mathbb{C}^{2^N}$.
We do not distinguish between operators on Hilbert space and matrices acting on $\mathbb{R}^{N_p}$; this is hopefully clear from context.
We use $\langle \cdot \rangle_{\rho}$ to denote an expectation value of an observable on a quantum state $\rho$, and $\overline{( \cdot )}$ to denote the expectation value of a random variable over all possible measurement outcomes.

\subsection{Hamiltonian learning as an oracle problem} \label{app:formal setup}

In this work, we consider a quantum experiment to consist of three steps: state preparation, unitary evolution, and measurement.
For a given experiment $x$, we denote the initial state as $\rho_x$.
We denote the unitary evolution by $U_x$.
Finally, a measurement is specified by a positive operator-valued measure (POVM), which corresponds to set of positive operators, $\{E_{x,m}\}$, that sum to the identity, $\sum_m E_{x,m} = \mathbb{1}$.
The measurement produces an outcome $m$ with probability,
\begin{equation}\label{eq:def_six}
    S_{x,m}=\mathrm{Trace}[E_{x,m} U_x \rho_ x U_x^\dagger].
\end{equation}
which is returned to the experimentalist.
In what follows, we will sometimes omit the experiment label $x$ when it is clear from context.

In this work, we are concerned with learning the parameters $\mathbf{u}$ of a Hamiltonian $H(\mathbf{u})$.
For specificity, we consider Hamiltonians that act on $N$ qubits and take the following form:
\begin{equation}\label{eq:H_no_control_formal}
    H(\mathbf{u})=\sum_{a=1}^{N_p} u_a P_a,\hspace{0.5cm} u_a\in[-1,1],\hspace{0.5cm} P_a\in\mathbb{P}^N,\hspace{0.5cm} P_a\neq I.
\end{equation}
where $u_a\in[-1,1]$ denotes an individual unknown parameter, $N_p$ is the total number of unknown parameters, and $P_a \in \{ \mathbb{1}, X, Y, Z \}^N$ is a Pauli operators on $N$ qubits. 
We are particularly interested in learning many-body Hamiltonians, when $N$ is large.

The data that we will learn $\mathbf{u}$ from is the outcomes of quantum experiments in which the unitary evolution $U_x$ involves time evolution by $H(\mathbf{u})$.
Before formally defining the Hamiltonian learning problem, we define three specific classes of such experiments that we will consider.
\begin{definition}[No quantum control]\label{def:no_quantum_control}
    A quantum experiment features \emph{no quantum control} if the unitary evolution is equal to a single application of time-evolution under the Hamiltonian,
    \begin{equation}\label{eq:U_no_control}
        U = e^{-i H(\mathbf{u}) t},
    \end{equation}
    for time $t$.
    Unless otherwise stated, we assume that state preparation and read-out can be performed in an arbitrary basis, and that the experimentalist can freely choose $t$ in each experiment.
\end{definition}
\noindent Experiments with no quantum control are the most restrictive class we will consider. 

We now consider two, idealized ways in which increased control by the experimentalist can enhance Hamiltonian learning experiments.
First, the experimentalist may be able to insert discrete instantaneous quantum gates throughout time-evolution~\cite{huang2022learning}:
\begin{definition}[Discrete quantum control]\label{def:discrete_quantum_control}
    A quantum experiment features discrete quantum control if the experimentalist can interleave time-evolution by a Hamiltonian $H(\mathbf{u})$ (Eq.~\ref{eq:H_no_control_formal}) with $L$ instantaneous unitary operations $V_l$. That is,
    \begin{equation}\label{eq:rhot_def_with_scattering}
    U = V_L e^{-iH\tau_L}V_{L-1}e^{-iH\tau_{L-1}}\ldots V_2e^{-iH\tau_2}V_1e^{-iH\tau_1}, \hspace{1cm} H = H(\mathbf{u})
    \end{equation}
    for some times $\tau_1,\ldots,\tau_L$.
    We assume that state preparation and read-out can be performed in an arbitrary basis, that the experimentalist can freely choose $\tau_l$ and $V_l$ in each experiment, and that $V_l$ may act on an arbitrary number of additional ancilla qubits.
\end{definition}
\noindent We note that the use of unitary control operations allows a large amount of generality.
For instance, an arbitrary mid-experiment POVM measurement can be implemented within the above definition as a unitary operation.
Moreover, our definition also incorporates experiments involving up to $L$ copies of a quantum memory, since the control operations can swap the qubits acted on by $H(\mathbf{u})$ with ancilla qubits.
Our consideration of a finite number $L$ of discrete operations is motivated by physical concerns, since any ``instantaneous'' operation will in practice require a non-zero amount of time and incur a non-zero amount of noise.
The above definition includes experiments with no quantum control as the special case $L=1$.
Also, the inclusion of the final unitary $V_L$ is not strictly necessary as it can be absorbed into the final measurement.

While the discrete quantum control model is natural for certain hybrid analog-digital quantum setups, in practice energetic restrictions often preclude the application of instantaneous control operations.
For example, when evolution under the Hamiltonian $H(\mathbf{u})$ is native to the system and cannot be paused, the application of instantaneous control operations requires driving the system with a control field orders of magnitude stronger than the native Hamiltonian strength.
To capture setups where only moderate control strength is possible, we consider the following alternate scenario.
We suppose that the experimentalist can \emph{continuously} modify time-evolution by augmenting the Hamiltonian with additional control parameters $\mathbf{c}$:
\begin{definition}[Continuous quantum control]\label{def:continuous_quantum_control}
    A quantum experiment features continuous quantum control if the unitary evolution is equal to a single application of time-evolution for time $t$ under an augmented Hamiltonian,
    \begin{equation}\label{eq:H_with_control_formal}
        U = e^{-i H(\mathbf{u},\mathbf{c}) t}, \,\,\,\,\,\,\, \mathrm{where} \,\,\,\,\,\,\, H(\mathbf{u},\mathbf{c})= H(\mathbf{u}) + H(\mathbf{c}) = \sum_{a=1}^{N_p} (u_{a}+c_a)P_a,\hspace{0.5cm} |u_a|,|c_a| \leq 1.
    \end{equation}
    Again, we assume that state preparation and read-out can be performed in an arbitrary basis, and that the experimentalist can choose $\mathbf{c}$ and $t$ in each experiment.
\end{definition}
\noindent We restrict the control parameters $\mathbf{c}$ to be time-independent, which turns out to be sufficient for our learning algorithm in Theorem~1 of the main text.
We also assume that the control parameters encompass the same Pauli operators as the unknown parameters.
This is an important and rather strict assumption that is necessary for our theorem.
Exploring Hamiltonian learning under more restrictive forms of continuous control (e.g.~when the control parameters are restricted to single-qubit operators) remains an interesting and potentially relevant future direction.

We can now formalize the Hamiltonian learning problem.
We do so by treating each quantum experiment as a query to a probabilistic classical oracle.
\begin{definition}[Classical oracularization of quantum experiments]\label{def:blackbox}
    A quantum experiment $x$ involving a Hamiltonian $H(\mathbf{u})$ with unknown parameters $\mathbf{u}$ takes as input a classical description of a quantum state $\rho_x$, a classical description of a POVM $\{E_{x,m}\}$, and: a time $t_x$ (if no quantum control), a list of $L$ times $\tau_{x,l}$, and $L$ unitary operations $V_{x,l}$ (if discrete quantum control), or a time $t_x$, and a list of control parameters $\mathbf{c}_x$ (if continuous quantum control).
    The oracle returns a measurement outcome $m$ drawn from the probability distribution $S_{x,m}(t)$ [Eq.~\eqref{eq:def_six}].
\end{definition}
\noindent The Hamiltonian learning problem is to learn $\mathbf{u}$ from calls to the classical oracle.
\begin{definition}[Hamiltonian learning problem]\label{def:learning_problem_formal}
    Consider a Hamiltonian $H(\mathbf{u})$ with unknown parameters $\mathbf{u}$, and a desired error $\epsilon$.
    The Hamiltonian learning problem is to construct an unbiased estimator $\tilde{\mathbf{u}}$ of $\mathbf{u}$ such that the maximum root-mean-square error is less than $\epsilon$,
    \begin{equation}
        \max_a \Big[\overline{(u_a-\tilde{u}_a)^2}\Big]^{1/2}\leq \epsilon,
    \end{equation}
    given access to an experiment oracle for $H(\mathbf{u})$. 
    We define the cost of a Hamiltonian learning algorithm as the total Hamiltonian evolution time, \begin{equation}
        T = \sum_x t_x, \,\,\,\,\,\,\,\,\, (\textnormal{where }  t_x = \sum_l\tau_{x,l} \textnormal{ for discrete quantum control})
    \end{equation}
    over all experiments performed.
\end{definition}

\subsection{Alternative error bounds}\label{app:alternativeerror}

In the above definition, we have chosen the maximum RMS error as a metric for the performance of a learning algorithm.
This is not the only choice possible, and different choices can lead to different costs to learn to the same $\epsilon$.
Popular performance metrics for the estimator $\tilde{\mathbf{u}}$ include:
\begin{enumerate}
    \item Have maximum RMS error $\max_a\sqrt{\overline{(u_a-\tilde{u}_a)^2}}\leq \epsilon$ (the metric used here).
    \item Have total RMS error
    $\sqrt{\sum_a\overline{(u_a-\tilde{u}_a)^2}}\leq \epsilon$.
    \item With probability $\geq 1-\delta$, satisfy $\|\mathbf{u}-\tilde{\mathbf{u}}\|_2\leq\epsilon$.
    \item With probability $\geq 1-\delta$, satisfy $\|\mathbf{u}-\tilde{\mathbf{u}}\|_{\infty}\leq\epsilon$.
\end{enumerate}
The RMS error $\sqrt{\overline{(u_a-\tilde{u}_a)^2}}$ is equivalent to the 2-norm of the standard deviation of the estimator $\tilde{u}_a$ under the assumption that this estimator is unbiased.
Thus, metric 2 is equal to the $2$-norm of the standard deviation of the vector estimator $\tilde{\mathbf{u}}$, and metric 1 is equivalent to the infinity norm of the standard deviation of $\tilde{\mathbf{u}}$.

Let us now outline the costs of converting between these various estimators.
We summarize this in Table~\ref{tab:conversion_summary}.
The contrapositive of a result ``existence of estimator $\tilde{\mathbf{u}}$ with cost T $\rightarrow$ existence of estimator $\tilde{\mathbf{u}}'$ with cost $T'$'' is the result ``a bound $T'$ on the cost of estimator $\tilde{\mathbf{u}}'$ $\rightarrow$ a bound $T$ on the cost of estimator $\tilde{\mathbf{u}}$'', so we will only discuss the costs of generating one estimator from another.
An estimator with $P(\|\tilde{\mathbf{u}}-\mathbf{u}\|_{\infty}\leq\epsilon)>1-\delta$ is an estimator with $P(\|\tilde{\mathbf{u}}-\mathbf{u}\|_{2}\leq N_p^{1/2}\epsilon)>1-\delta$, and an estimator with $P(\|\tilde{\mathbf{u}}-\mathbf{u}\|_{2}\leq\epsilon)>1-\delta$ is an estimator with $P(\|\tilde{\mathbf{u}}-\mathbf{u}\|_{\infty}\leq\epsilon)>1-\delta$
as 
\begin{equation}
    N_p^{-\frac{1}{2}}\|\mathbf{u}-\tilde{\mathbf{u}}\|_2\leq\|\mathbf{u}-\tilde{\mathbf{u}}\|_{\infty}\leq\|\mathbf{u}-\tilde{\mathbf{u}}\|_{2}.
\end{equation}
Similar logic holds for the RMS error, as 
\begin{equation}
    \max_a \overline{(u_a-\tilde{u}_a)^2}\leq \sum_a\overline{(u_a-\tilde{u}_a)^2}\leq N_p\max_a \overline{(u_a-\tilde{u}_a)^2}.
\end{equation}
As we have not declared for these estimators how $T$ scales with $\epsilon$, we cannot propagate these changes through to a change in $T$.

We now consider how to change from bounds on the RMS error to probability-confidence bounds and vice versa.
Given an estimator $\tilde{\mathbf{u}}$ with total RMS error $\epsilon$, Chernoff's bound implies we can take the mean of $\mathcal{O}(\log(1/\delta))$ repeated estimates to guarantee $P(\|\mathbf{u}-\tilde{\mathbf{u}}\|_2\leq\epsilon)\geq 1-\delta$.
Conversely, given an estimator $\tilde{\mathbf{u}}$ with $P(\|\mathbf{u}-\tilde{\mathbf{u}}\|_2\leq\epsilon)\geq 1-\delta$, as $\|\mathbf{u}-\tilde{\mathbf{u}}\|_2\leq N_p^{1/2}$, we can take a median over $\mathcal{O}(\log(N_p)\log(1/\epsilon)/\log(1/\delta))$ estimation copies to obtain an estimator with total RMS error $2\epsilon$.
Similarly, given an estimator $\tilde{\mathbf{u}}$ with $P(\|\mathbf{u}-\tilde{\mathbf{u}}\|_\infty\leq\epsilon)\geq 1-\delta$, I can take a term-wise median over $\mathcal{O}(\log(1/\epsilon)/\log(1/\delta))$ estimation copies to obtain an error with maximum RMS error $\epsilon$.
Then, given an estimator $\tilde{\mathbf{u}}$ with maximum RMS error $\epsilon$, we can take a mean over $\mathcal{O}(\log(1/\delta))$ copies and for each $a$ construct an estimator $\tilde{u}_a$ satisfying $P(|u_a-\tilde{u}_a|\leq\epsilon)>1-\delta$.
But, these estimators can be correlated; in the worst case we can only bound $P(\max|u_a-\tilde{u}_a|\geq\epsilon) \leq \min(1, N\delta)$.
To achieve $P(\|\mathbf{u}-\tilde{\mathbf{u}}\|_{\infty}\leq\epsilon)\leq 1-\delta'$ then requires we set $\delta=\delta'/N$, which incurs an additional $\mathcal{O}(\log(N))$ cost.
All other conversions in Tab.~\ref{tab:conversion_summary} can be obtained by propagating through the results outlined above.
One can see that our chosen metric --- the max RMS error --- is relatively weak (in that it is typically more costly to expand to another metric than vice-versa).
This implies that algorithms to achieve a given~$\epsilon$ are easier to construct, but bounding the cost of an algorithm to achieve accuracy $\epsilon$ is more challenging.

\begin{table}[]
    \centering
    \scriptsize
    \bgroup
    \setlength{\tabcolsep}{1pt}
    \def\arraystretch{1.5}
    \begin{tabular}{|x{0.2\linewidth}|x{0.22\linewidth}|x{0.2\linewidth}|x{0.2\linewidth}|x{0.2\linewidth}|}\hline
        $T'\downarrow\hspace{1cm} T\rightarrow$
        & $\max_a[\overline{(u_a-\tilde{u}_a)^2}]^{1/2}\leq \epsilon$ 
        & $[\sum_a\overline{(u_a-\tilde{u}_a)^2}]^{1/2}\leq \epsilon$ 
        & $P(\|\tilde{\mathbf{u}}-\mathbf{u}\|_{2}\leq\epsilon)>1-\delta$
        & $P(\|\tilde{\mathbf{u}}-\mathbf{u}\|_{\infty}\leq\epsilon)>1-\delta$
        \\\hline
        $\max_a[\overline{(u_a-\tilde{u}_a)^2}]^{1/2}\leq \epsilon'$ 
        & -
        & $\epsilon' = \epsilon$,
        & $\epsilon' = \epsilon$,
        & $\epsilon' = \epsilon$,
        \\
        &
        & $T' = T$
        & $T' = \mathcal{O}(T\frac{\log(1/\epsilon)}{\log(1/\delta)})$
        & $T' = \mathcal{O}(T\frac{\log(1/\epsilon)}{\log(1/\delta)})$
        \\\hline
        $[\sum_a\overline{(u_a-\tilde{u}_a)^2}]^{1/2}\leq \epsilon'$
        & $\epsilon' = N_p^{1/2}\epsilon$,
        & -
        & $\epsilon' = \epsilon$
        & $\epsilon'=N_p^{1/2}\epsilon$,
        \\
        &  $T'=T$
        &
        & $T' = \mathcal{O}(T\frac{\log(N_p)\log(1/\epsilon)}{\log(1/\delta)})$
        & $T' = \mathcal{O}(T\frac{\log(N_p)\log(1/\epsilon)}{\log(1/\delta)})$
        \\\hline
        $P(\|\tilde{\mathbf{u}}-\mathbf{u}\|_{2}\leq\epsilon')>1-\delta'$
        & $\epsilon'=N_p^{1/2}\epsilon$,
        & $\epsilon'=\epsilon$,
        & -
        & $\epsilon' = N_p^{1/2}\epsilon$,
        \\
        & $T' = \mathcal{O}(T\log(N_p)\log(1/\delta'))$
        & $T'=\mathcal{O}(T\log(1/\delta'))$
        &
        &  $T'=T$, $\delta'=\delta$
        \\\hline
        $P(\|\tilde{\mathbf{u}}-\mathbf{u}\|_{\infty}\leq\epsilon')>1-\delta'$
        & $\epsilon' = \epsilon$
        & $\epsilon' = \epsilon$
        & $\epsilon'=\epsilon$,
        & -
        \\
        & $T' = \mathcal{O}(T\log(N_p)\log(1/\delta'))$
        & $T'=\mathcal{O}(T\log(1/\delta'))$
        &  $T'=T$, $\delta'=\delta$
        &
        \\\hline
    \end{tabular}
    \egroup
    \caption{Conversion table between well-known estimators. Entry $i,j$ gives the conversion from estimator $j$ (defined in top row) to estimator $i$ (defined in left-column). Equivalently, entry $i,j$ gives the conversion from a bound on estimator $i$ (defined in left column) to a bound on estimator $j$ (defined in top row).
    As indicated in the top-left corner, we assume the cost of constructing the estimator for column $j$ is $T$, and the cost of constructing the estimator for row $i$ is $T'$.}
    \label{tab:conversion_summary}
\end{table}

\subsection{The Fisher information and the Cramer-Rao bound}\label{app:fisher_information}

The ability to infer the hidden parameters $\mathbf{u}$ from the experimental outcomes $\{m_x\}$ is bounded by how sensitive the probability distributions $S_{x,m}$ are to changes in $\mathbf{u}$.
The celebrated Cramer-Rao theorem solidifies this, by bounding the covariance matrix of any unbiased estimation of $\mathbf{u}$ by the inverse of the Fisher information matrix $\bs{\mathcal{I}}$~\cite{liu2017quantum}.
The Fisher information matrix quantifies the sensitivity of $S_{x,m}$ to $\mathbf{u}$, and has matrix elements:
\begin{align}
    \mathcal{I}_{ab}
    =\sum_x \mathcal{I}_{x,ab}&=\sum_{m,x}S_{x,m}\Bigg(\frac{\partial \log S_{x,m}}{\partial u_a}\Bigg)
    \Bigg(\frac{\partial \log S_{x,m}}{\partial u_b}\Bigg) =\sum_{m,x} \frac{1}{S_{x,m}}
    \Bigg(\frac{\partial S_{x,m}}{\partial u_a}\Bigg)
    \Bigg(\frac{\partial S_{x,m}}{\partial u_b}\Bigg).
\end{align}
where $a,b = 1,\ldots,N_p$ index the unknown parameters. 
We have used the fact that the Fisher information is additive to write it as a sum over experiments $x$, and then differentiated the logarithm.
In the experiments considered in this work, the dependence of $S_{x,m}$ on $\mathbf{u}$ arises from time-evolution under the unknown Hamiltonian within $U_x$.
The Cramer-Rao theorem establishes that this is the maximum amount of information that can be obtained from the set of experiments $\{x\}$ with probabilistic outcomes $\{ m_x \}$, regardless of whether the experiments were performed adaptively or not.

The classical Fisher information above depends upon the measurement bases $E_{m,x}$ of each experiment $x$.
One can also ask the question: what is the maximum information extractable from the final state $U_x \rho_x U_x^\dag$ of each experiment by any possible measurement.
This is quantified by the quantum Fisher information~\cite{braunstein1994statistical,braunstein1996generalized}.
For the experiments considered in this work, we assume the initial state $\rho_x$ is pure and the dependence on the unknown parameters enters only through the unitary evolution $U_x$.
The quantum Fisher information matrix then takes the simple form~\cite{liu2020quantum}:
\begin{equation}\label{eq:QFI_pure_states}
    \mathcal{I}_{x,ab}^{(Q)} =4 \left( \expval{A_{x,a} A_{x,b}}_{\rho_x} - \expval{A_{x,a}}_{\rho_x} \expval{A_{x,b}}_{\rho_x} \right),
\end{equation}
where we denote $\expval{\cdot}_\rho \equiv \mathrm{Trace}[ (\cdot) \rho]$.
(Note that maximal quantum Fisher information from a unitary process is always achieved for a pure state, and so the bounds we derive in this work will hold for mixed states $\rho_x$ as well~\cite{liu2020quantum}.)
Here, we define a central object for learning unitary quantum processes, the Hermitian time-integrated perturbation operator:
\begin{equation}\label{eq:time_integrated_perturbation_def}
    A_{x,a}=-iU_x^{\dag} \frac{\partial U_x}{\partial u_a}.
\end{equation}
For convenience, we write the same equations in vector notation as:
\begin{equation} \label{I A definition vector}
    \bs{\mathcal{I}}^{(Q)}_x = 4 \left( \expval{ \bs{A}_x \bs{A}_x^T}_{\rho_x} - \expval{\bs{A}_x}_{\rho_x} \expval{\bs{A}^T_x}_{\rho_x} \right), \,\,\,\,\,\,\,\,\,\, \bs{A}_x = -iU^\dagger_x \left( \bs{\nabla} U_x \right),
\end{equation}
where the transpose $T$ acts in parameter space (not  Hilbert space).

The quantum Cramer-Rao theorem bounds the covariance matrix of any unbiased estimation of $\mathbf{u}$ in terms of the quantum Fisher information matrix, regardless of the choice of measurement bases $E_{m,x}$.
We state it without proof:
\begin{theorem}[Quantum Cramer-Rao theorem \cite{liu2020quantum}] \label{thm:cramerrao}
Let $\{m_x\}$ be measurement outcomes of a set of quantum experiments $x$ whose unitary evolution depends on an unknown parameter $\mathbf{u}$, and define the quantum Fisher information matrix $\bs{\mathcal{I}}^{(Q)} = \sum_x \bs{\mathcal{I}}^{(Q)}_x$ following Eq.~(\ref{eq:QFI_pure_states}).
Suppose that $\tilde{\mathbf{u}}(\{ m_x \})$ is an unbiased estimator of $\mathbf{u}$ from the outcomes $\{m_x\}$.
Then the covariance matrix of the estimator,
\begin{equation}
    \bs{\Sigma} = \overline{ \left( \tilde{\mathbf{u}} - \mathbf{u} \right) \left( \tilde{\mathbf{u}} - \mathbf{u} \right)^T}
\end{equation}
is lower bounded by the inverse quantum Fisher information matrix:
\begin{equation}
    \bs{\Sigma} \geq [\bs{\mathcal{I}}^{(Q)}]^{-1}.
\end{equation}
Here $\overline{( \cdot )}$ denotes the expectation value over measurement outcomes $\{ m_x \}$.
\end{theorem}
\noindent The quantum Cramer-Rao theorem immediately allows us to lower bound the error metrics introduced previously.
The root-mean-square error in an individual parameter $u_a$ is lower bounded by the diagonal elements of the inverse Fisher information matrix:
\begin{equation}
    \overline{ \left( u_a - \tilde{u}_a \right)^2 } \geq [\bs{\mathcal{I}}^{(Q)}]^{-1}_{aa},
\end{equation}
and the total root-mean-square error by its trace:
\begin{equation}
    \sum_a \overline{ \left( u_a - \tilde{u}_a \right)^2 } \geq \mathrm{Trace}( [\bs{\mathcal{I}}^{(Q)}]^{-1} ).
\end{equation}
Following the previous section, we note that the trace also lower bounds our error metric of choice, the maximum individual root-mean-square error $\epsilon$, via:
\begin{equation} \label{eq:our_CR_bound}
    \epsilon 
    = \max_a \left( \overline{ \left( u_a - \tilde{u}_a \right)^2 } \right)^{1/2} 
    \geq  \left( \max_a [\bs{\mathcal{I}}^{(Q)}]^{-1}_{aa} \right)^{1/2} 
    \geq \left( \frac{1}{N_p} \mathrm{Trace}( [\bs{\mathcal{I}}^{(Q)}]^{-1} ) \right)^{1/2}.
\end{equation}

We are now able to formally derive the Heisenberg limit for Hamiltonian learning:
\begin{theorem}[Heisenberg limit in Hamiltonian learning]\label{thm:heisenberg}
Consider a set of quantum experiments $x$ with unitary evolution, 
\begin{equation}
U_x = \mathcal{T} \left\{ e^{-i \int_0^{t_x} ds \, H_x(s,\mathbf{u})} \right\},
\end{equation}
where $H_x(t,\mathbf{u})$ is an arbitrary time-dependent Hamiltonian with a bounded dependence on the unknown parameters $\mathbf{u}$, 
\begin{equation}
\left\lVert \frac{\partial H_x(s,\mathbf{u})}{\partial u_a} \right\rVert_s \leq 1, \,\,\,\,\, \forall \, a.
\end{equation}
Then any algorithm that solves the Hamiltonian learning problem requires a total evolution time,
\begin{equation}
    T \geq \Omega(\epsilon^{-1}),
\end{equation}
where $T = \sum_x t_x$.
\end{theorem}
\noindent In the above, we allow a completely general dependence of the unitary evolution on $\mathbf{u}$ in order to capture both the discrete and continuous quantum control scenarios.

\emph{Proof---}
Using the definition of the parametric derivative of an operator exponential \cite{wilcox1967exponential}, we have
\begin{equation}
   \frac{\partial U_x}{\partial u_a} = -i\int_0^{t_x}ds \, U_x(t_x,s)\frac{\partial H_x(s,\mathbf{u})}{\partial u_a} U_x(s,0),
\end{equation}
where $U_x(t,s)$ denotes evolution from time $s$ to time $t$.
Observing Eq.~(\ref{eq:time_integrated_perturbation_def}), we can then write the time-integrated perturbation vector as follows:
\begin{equation}\label{eq:generator_integral_form}
    A_{x,a}=- \int_0^{t_x}ds \, U_x^\dagger(s,0)\frac{\partial H_x(s,\mathbf{u})}{\partial u_a} U_x(s,0).
\end{equation}
We can bound the spectral norm as $\lVert A_{x,a} \rVert_s \leq t_x$, using the triangle inequality and the assumption $\lVert \frac{\partial H_x(s,\mathbf{u})}{\partial u_a} \rVert_s \leq 1$.
This bounds the diagonal elements of the inverse quantum Fisher information matrix as,
\begin{equation}\label{eq:QFI_bound}
[\bs{\mathcal{I}}^{(Q)}]^{-1}_{aa} 
\geq \frac{1}{\mathcal{I}^{(Q)}_{aa}}
\geq \frac{1}{4 \sum_x t_x^2} \geq \frac{1}{4 T^2}.
\end{equation}
Applying the quantum Cramer-Rao bound (Theorem~\ref{thm:cramerrao}) and Eq.~(\ref{eq:our_CR_bound}) yields,
\begin{equation}
    T \geq \frac{1}{2 \epsilon}
\end{equation}
which is the Heisenberg limit. \qed

\section{A Heisenberg-limited algorithm for low-intersection Hamiltonians with continuous quantum control}\label{app:HKT_Heisenberg}

\subsection{The Haah-Kothari-Tang algorithm for learning from short time evolution}\label{app:HKT}
In Ref.~\cite{haah2021optimal}, Haah, Kothari and Tang described an algorithm to learn a Hamiltonian $H$ from short-time evolution by $e^{-iHt}$, along with state preparation and measurement in an arbitrary product basis.
We restate their result here in our formalism for ease of reading.

The HKT algorithm for finite times requires taking a Taylor expansion of $e^{-iHt}Pe^{iHt}$, which in turn requires that this expansion converges. This sets a maximum time, $t_c$, beyond which the algorithm no longer works.
Ref.~\cite{haah2021optimal} calculates $t_c$, in the case that $P_a$ is a Pauli operator with support on $\mathcal{O}(1)$ qubits, as follows.
Let us define $\mathrm{Supp(P)}$ as the set of qubits that $P$ acts non-trivially on, and let us further define
\begin{equation}\label{eq:frakd_def}
     \mathfrak{d}=\max_{a=1,\ldots, N_p}\Big|\Big\{b=1,\ldots, N_p;\, b\neq a;\,\mathrm{Supp}(P_a)\cap\mathrm{Supp}(P_b)\neq\emptyset\Big\}\Big|,
\end{equation}
which is the maximum degree in the dual graph to $H$. Ref.~\cite{haah2021optimal} gives a critical time $t_c=\mathcal{O}(\mathrm{Poly}(\mathfrak{d}^{-1}))$.
Note that if $H$ has maximum degree $d-1=\max_q|\{a=1,\ldots,N_p;\, q\in\mathrm{Supp}(P_a)\}|$, we can bound $d\leq \mathfrak{d}\leq k(d-1)$.
It is typically assumed that $\mathfrak{d}$ is constant in $N$, in which case the Hamiltonian is called `low-intersection', but we will not use this terminology in this work.
With this fixed, Ref.~\cite{haah2021optimal} gives an algorithm (the HKT algorithm) for learning at finite times.
We refer the reader to the original work for details of the algorithm, and state it in a form relevant to us here:
\begin{theorem}[\cite{haah2021optimal} A.1, restated]\label{thm:HKT_algorithm}
    Consider the Hamiltonian learning problem for an unknown Hamiltonian $H(\mathbf{u})$, with $\lVert \mathbf{u} \rVert_\infty < 1$, with no quantum control (Definition~\ref{def:no_quantum_control}).
    There exists an algorithm (the HKT algorithm) that with probability $1-\delta$ can estimate $\mathbf{u}$ to an error $\|\tilde{\mathbf{u}}-\mathbf{u}\|_{\infty}\leq\epsilon$ in total experimental time,
    \begin{equation}\label{eq:HKT_runtime}
    T=\mathcal{O}\bigg(\frac{\mathrm{Poly}(\mathfrak{d})}{\epsilon^2}\log\frac{N_p}{\delta}\bigg),
    \end{equation}
    where $N_p$ is the number of parameters, and $\mathfrak{d}$ is the maximum degree in the dual graph [Eq.~\eqref{eq:frakd_def}].
\end{theorem}
\noindent
The HKT algorithm does not achieve the Heisenberg limit, $T = \mathcal{O}(\epsilon^{-1})$, since each experiment involves time-evolution only for times $t\leq t_c$.

\subsection{A Heisenberg-limited extension of the HKT algorithm using continuous quantum control}

In this section, we extend the HKT algorithm described above to a new algorithm that works at the Heisenberg limit, and prove Theorem~1 of the main text.
The idea is, as we improve our estimate of $\mathbf{u}$, to add a control term $\mathbf{c}\sim-\mathbf{u}$, so that the terms in Eq.~\eqref{eq:H_with_control_formal} approximately cancel out.
We can rescale the resulting Hamiltonian, allowing us to `zoom in' on the error between our control and fixed parameters at the Heisenberg limit.
This is similar in spirit to robust phase estimation~\cite{kimmel2015robust} (and its multiple phase extension~\cite{dutkiewicz2022heisenberg}), which achieves the Heisenberg limit by using estimates of $e^{i2^d\phi}$ to estimate $\phi$ to within an error of $\frac{\pi}{2^{d-1}}$.
Doing this for $d=0,1,\ldots D$ avoids the aliasing error $e^{i2^d(\phi+2^{d-1}\pi)}=e^{i2^d\phi}$.

To achieve the Heisenberg limit requires we assume continuous quantum control of our system (Definition~\ref{def:continuous_quantum_control}).
This allows us to `zoom in' on our Hamiltonian: we use our control term to approximately cancel our fixed parameters, and rescale our time variable.
This gives a new black box that we can plug into the HKT algorithm, but one which takes longer to query.
\begin{lemma}\label{lem:black_box_rescaling}
    Assume we have an experiment oracle time-independent $H(\mathbf{u},\mathbf{c})$ with dynamic control, and an estimate $\tilde{\mathbf{u}}\in [-1,1]^{N_p}$, such that $\|\mathbf{u}-\tilde{\mathbf{u}}\|_{\infty} < \Delta$.
    We can query a `rescaled' experiment oracle with a Hamiltonian $H'(\mathbf{u}')$, where $\mathbf{u}'=\frac{1}{\Delta}(\mathbf{u}-\tilde{\mathbf{u}})$ and $\lVert u'\rVert_\infty \leq 1$.
    This incurs a rescaled cost $\frac{t}{\Delta}$ to query for an experiment with evolution time $t$.
\end{lemma}
\emph{Proof---}If $\|\mathbf{u}-\mathbf{c}\|_{\infty}\leq\Delta$, $\|\Delta^{-1}(\mathbf{u}-\mathbf{c})\|_{\infty}\leq 1$ and $\Delta^{-1}(\mathbf{u}-\mathbf{c})\in [-1,1]^{N_p}$.
Then, fix input $\rho(0),\mathbf{0},t,\{E_{i}\}$, for the new black box.
From Eq.~\eqref{eq:H_with_control_formal} we can see that $H(\mathbf{a},\mathbf{b})=H(\mathbf{a}+\mathbf{b},0)$ and $H(\lambda\mathbf{a},0)=\lambda H(\mathbf{a},0)$, which implies we need to sample from
\begin{equation}
S_{m}(t)=\mathrm{Trace}\Big[E_{m}\exp\big(-it\Delta^{-1}H(\mathbf{u},-\mathbf{c},0)\big)\rho(0)\exp\big(it\Delta^{-1}H(\mathbf{u},-\mathbf{c})\big)\Big].
\end{equation}
We can obtain this sampling by querying the original black box with $\rho=\rho(0),\mathbf{c}=-\tilde{\mathbf{u}},t=t/\Delta,\{E_{i}\}=\{E_{i}\}$, which has a cost $\frac{t}{\Delta}$.\qed 

With this given, our algorithm proceeds as follows:
\begin{algorithm}[Heisenberg-limited extension of the HKT algorithm]\label{alg:HHKT}
Input: Hamiltonian $H=H(\mathbf{u},\mathbf{c})$ as given by Eq.~\eqref{eq:H_with_control_formal}, desired precision $\epsilon$, confidence parameter $c \in (0,1/24]$.
\begin{enumerate}
    \item Let $D = \lceil\log_2(1/\epsilon)\rceil$.
    \label{step:def_D}
    \item Let $\tilde{\mathbf{u}}^{(0)} = \mathbf{0}$ be our $0$th order estimate for $\mathbf{u}$.
    \item For $d = 0, ..., D$:
    \begin{enumerate}
        \item Fix $H^{(d)}(\mathbf{u}) = 2^d H(\mathbf{u},-\tilde{\mathbf{u}}^{(d)})$.
        \item Apply the HKT algorithm (Theorem~\ref{thm:HKT_algorithm}) to the rescaled Hamiltonian $H^{(d)}$, with error bound $1/2$ and error probability $\delta^{(d)} = \frac{c}{8^{D-d}}$.
        Define the output estimate ${\mathbf{g}}^{(d)}$.
        \label{step:HKT_sub}
        \item If $\|\mathbf{g}^{(d)}\|_{\infty}\geq 1$, set $\mathbf{g}^{(d)}=0$.
        \label{step:bound_norm_g}
        \item Set $\tilde{\mathbf{u}}^{(d+1)}=\tilde{\mathbf{u}}^{(d)}+2^{-d}\mathbf{g}^{(d)}$.
        \label{step:new_estimate}
    \end{enumerate}
    \item Return $\tilde{\mathbf{u}} =\tilde{\mathbf{u}}^{(D+1)}$. 
\end{enumerate}
\end{algorithm}

To show that this algorithm achieves the Heisenberg limit as defined, we first bound the error in the $d$-th order estimate $\mathbf{\tilde u}^{(d)}$ assuming the first $d$ runs of the HKT subroutine have succeeded (Lemma~\ref{lem:bounding_faultless_HHKT_performance}).
We then bound the error in the final estimate $\mathbf{\tilde u}$ under the same assumption (Lemma~\ref{lem:bounding_fault_HHKT_performance}).
Finally, in Theorem~\ref{thm:HHKT_performance}, we combine these results to show the final RMS error is $\leq \epsilon$, and calculate the cost of running Algorithm~\ref{alg:HHKT} to prove the Heisenberg scaling.

\begin{lemma}\label{lem:bounding_faultless_HHKT_performance}
    Consider an application of Algorithm~\ref{alg:HHKT} to an experiment oracle with unknown parameters $\mathbf{u}$. If for each $d'=0,1,\ldots,d-1$ the HKT algorithm subroutine in step \ref{step:HKT_sub} of the algorithm succeeds, then $\|\tilde{\mathbf{u}}^{(d)}-\mathbf{u}\|_\infty\leq 2^{-d}$. 
\end{lemma}
\emph{Proof---}We prove this result by induction in $d$. 
First we prove the case for $d=1$.
As $H^{(1)}(\mathbf{u}) = H(\mathbf{u})$ we always satisfy the input condition of the HKT algorithm $\lVert \mathbf{u} \rVert_\infty \leq 1$. We have $\tilde{\mathbf{u}}^{(1)} = g^{(1)}$, hence the success of the HKT algorithm implies $\|\tilde{\mathbf{u}}^{(1)}-\mathbf{u}\|\leq 1/2$.

Next, we prove that $d\rightarrow d+1$.
Lemma~\ref{lem:black_box_rescaling} implies that if $\|\tilde{\mathbf{u}}^{(d)}-\mathbf{u}\|\leq 2^{-d}$ we satisfy the input conditions to the HKT algorithm at round $d$.
If the subroutine succeeds, the output will satisfy
\begin{equation}
\|{\mathbf{g}}^{(d)} - 2^d(\mathbf{u}-\tilde{\mathbf{u}}^{(d)})\|\leq\frac{1}{2}.
\end{equation}
Multiplying both sides of this inequality by $2^{-d}$, and substituting in the definition of $\tilde{\mathbf{u}}^{(d+1)}$ we find that $\|\tilde{\mathbf{u}}^{(d+1)}-\mathbf{u}\|\leq 2^{-(d+1)}$, as required.\qed

\begin{lemma}\label{lem:bounding_fault_HHKT_performance}
    Consider an application of Algorithm~\ref{alg:HHKT} to an experiment oracle with unknown parameters $\mathbf{u}$. If for each $d'=0,1,\ldots,d-1$ the HKT algorithm subroutine in step \ref{step:HKT_sub} of the algorithm succeeds, then the final estimate $\tilde{\mathbf{u}}$ satisfies $\|\tilde{\mathbf{u}}-\mathbf {u}\|_\infty\leq 3\times2^{-d}$. 
\end{lemma}
\emph{Proof---}
We will first show that the final estimate $\mathbf{\tilde u}$ is close to the $d$-th order estimate $\mathbf{\tilde u}^{(d)}$. Combining this observation with Lemma~\ref{lem:bounding_faultless_HHKT_performance} completes the proof.

We have defined the estimate $\mathbf{\tilde u} = \mathbf{\tilde u}^{(D+1)}$ recursively in step \ref{step:new_estimate} of Algorithm \ref{alg:HHKT}. We can expand this as
\begin{equation}
     \mathbf{\tilde u} = 
     \mathbf{\tilde u} ^{(D)}+  \frac{\mathbf{g}^{(D)}}{2^{D}}
     =
     \mathbf{\tilde u} ^{(D-1)}+ \frac{\mathbf{g}^{(D-1)}}{2^{D-1}}+  \frac{\mathbf{g}^{(D)}}{2^{D}} = ... = 
     \mathbf{\tilde u} ^{(d)}+ \sum_{d'=d}^{D} \frac{\mathbf{g}^{(d')}}{2^{d'}}.
\end{equation}
As in step \ref{step:bound_norm_g} we ensure $\lVert \mathbf{g}^{(d')} \rVert_\infty \leq 1$ for all $d'$, by the triangle inequality we have
\begin{equation}
    \lVert \mathbf{\tilde u} - \mathbf{\tilde u} ^{(d)}  \rVert_\infty \leq \sum_{d'=d}^{D} 2^{-d'}\lVert \mathbf{g}^{(d')} \rVert_\infty \leq 2^{-d+1}.
    \label{eq:estimates_diff_bound}
\end{equation}

We use the triangle inequality again to bound the error in the final estimate
\begin{equation}
     \lVert \mathbf{\tilde u} - \mathbf{u} \rVert_\infty \leq  \lVert \mathbf{\tilde u} - \mathbf{\tilde u} ^{(d)}  \rVert_\infty +  \lVert \mathbf{\tilde u} ^{(d)} - \mathbf{u} \rVert_\infty.
\end{equation}
Bounding the first term as in Eq.~\eqref{eq:estimates_diff_bound}, and the second term as in Lemma~\ref{lem:bounding_faultless_HHKT_performance} yields the required result.
\qed

Now, let us calculate the cost of implementing Algorithm~\ref{alg:HHKT} and show that it scales at the Heisenberg limit.
\begin{theorem}\label{thm:HHKT_performance}
    Given access to an experiment oracle with continuous quantum control, Algorithm~\ref{alg:HHKT} solves the learning problem (Definition~\ref{def:learning_problem_formal}) to RMS error $\epsilon$ at a cost $T = \mathcal{O}(\epsilon^{-1})$.
\end{theorem}
\emph{Proof---}We first show that the RMS error in the final estimate $\mathbf{\tilde u}$ is $\leq 2^{-D} \leq \epsilon$, where the second inequality follows from the definition of the final order $D$ in step \ref{step:def_D}.
We have to consider contributions to the error from two different scenarios: either the HKT subroutine in step \ref{step:HKT_sub} succeeds every time, or it fails for the first time at some order $d$.
If the subroutine always succeeds, which happens with probability $\prod_{d'=0}^{D} (1-\delta^{(d')})\leq 1$, by Lemma~\ref{lem:bounding_faultless_HHKT_performance} the final error is at most $2^{-(D+1)}$.
With probability $\delta^{(d)}\prod_{d'=0}^{d-1} (1-\delta^{(d')})\leq \delta^{(d)} =c/8^{D-d}$, the HKT subroutine succeeds for $d'=0,1,\dots,d-1$ and fails at order $d$. Lemma~\ref{lem:bounding_faultless_HHKT_performance} bound the final error in this case by $3\times 2^{-d}$.
Combining all the possible cases, we can bound the squared error as
\begin{align}
    \overline{(\tilde u_a - u_a)^2}
    &\leq 2^{-2(D+1)} + \sum_{d = 0}^{D}c/8^{D-d}\times3^2\times2^{-2d}\\
    & = 2^{-2(D+1)} + 9c \sum_{d = 0}^{D}2^{-3D+d}.
\end{align}
Summing the geometric series in the second term we get
\begin{align}
    \overline{(\tilde u_a - u_a)^2}
    &\leq 2^{-2D} \left(\frac{1}{4}+ 18c \right).
\end{align}
As we have assumed $c \leq 1/24$, the right hand side is $\leq 2^{-2D} \leq \epsilon^2$. The above equation holds for all $a$, which implies the desired bound on the maximum RMS error.

It remains to calculate the cost of executing the algorithm
Let us write this as a sum $C=\sum_d C_d$ where $C_d$ is the cost of the $d$th order call to the HKT algorithm
Substituting in $\epsilon=1/2$ and $p=1-c/8^{D-d}$, and recalling that the runtime of the $d$th order black-box experiment scales as $2^d$, we have
\begin{equation}
C_d = \mathcal{O}\Bigg(\mathrm{Poly}(\mathfrak{d})2^d\log\frac{8^{D-d} N}{c}\Bigg).
\end{equation}
We can then evaluate $C=\sum_dC_d$ by summing the geometric series,
\begin{align}
\sum_d \mathrm{Poly}(\mathfrak{d}) 2^d\log\frac{8^{D-d} N}{c}&=\mathrm{Poly}(\mathfrak{d})\sum_{d=0}^{D}\Big(\log(N/c) + (D-d)\log(8)\Big) 2^d\\&=\mathrm{Poly}(\mathfrak{d})\Bigg[\log(N/c) \Big(2^{D+1}-1 \Big)+\log(8)\Big(2^{D+1}-D-2\Big)\Bigg],
\end{align}
and keeping only terms at leading order in $2^D$, yields,
\begin{equation}
C = \sum_d C_d = \Theta\Bigg(2^D\mathrm{Poly}(\mathfrak{d})\log\frac{8N}{c}\Bigg) = \Theta\Bigg(\frac{\mathrm{Poly}(\mathfrak{d})\log\frac{8N}{c}}{\epsilon}\Bigg).
\end{equation}
We have $C = \Theta(\epsilon^{-1})$as required.\qed

\section{Absence of the Heisenberg limit under unitary invariance}\label{app:no_control_no_heisenberg}

In this section, we establish our first no-go theorem, showing that learning at the Heisenberg limit is  impossible without quantum control, for certain Hamiltonians. We in fact establish a moderately stronger result: that learning at the Heisenberg limit is  impossible even with discrete quantum control, unless the number $L$ of interleaved unitaries scales with the total time $T$, $L = \Omega(T)$.
The key result underpinning this section is a bound on the quantum Fisher information matrix as a function of the derivatives of eigenvectors and eigenvalues of the Hamiltonian.
We state this for the discrete quantum control model with fixed $L$, and recall that this is equivalent to  no quantum control when $L=1$.

We begin by relating the quantum Fisher information to the eigenvalue decomposition of the Hamiltonian through the following Lemma.
\begin{lemma}\label{lem:Hamiltonian_derivative_bound}
    Consider a quantum experiment involving a Hamiltonian, $H(\mathbf{u})$, in the discrete quantum control model with up to $L$ interleaved unitaries.
    Write the eigenvalue decomposition of the Hamiltonian as $H(\mathbf{u})=W(\mathbf{u})^{\dag} D(\mathbf{u}) W(\mathbf{u})$, where $W$ is unitary and $D$ is diagonal.
    Then for any normalized $\mathbf{v} \in \mathbb{R}^{N_p}$, we have:
    \begin{equation}
    \mathbf{v}^T\bs{\mathcal{I}}^{(Q)}\mathbf{v} \leq \min \bigg( t\|\partial_{\mathbf{v}}D\|_s + 2L\|\partial_{\mathbf{v}} W\|_s\,\, , \,\, t\|\partial_{\mathbf{v}}H\|_s\bigg)^2,
    \end{equation}
    where $\bs{\mathcal{I}}^{(Q)}$ is the quantum Fisher information matrix for the experiment and $\partial_{\mathbf{v}} \equiv \sum_a v_a\frac{\partial}{\partial u_a}$.
\end{lemma}
\emph{Proof---}
Observing the definition of the quantum Fisher information matrix [Eq.~\eqref{eq:QFI_pure_states}], we immediately have
\begin{equation} \label{eq: I less than A2}
\mathbf{v}^T\bs{\mathcal{I}}^{(Q)}\mathbf{v} = \langle A_\mathbf{v}^2\rangle -\langle A_\mathbf{v}\rangle^2 \leq \langle A_{\mathbf{v}}^2\rangle \leq \|A_{\mathbf{v}}\|_s^2.
\end{equation}
Here we denote $A_\mathbf{v} = \sum_a v_a A_a$, with the operator $A_a$ defined in terms of the time-evolution unitary in Eq.~(\ref{eq:time_integrated_perturbation_def}), $A_a = U(t,0)^\dagger \frac{\partial}{\partial u_a} U(t,0)$.
Working in the discrete quantum control model [Eq.~\eqref{eq:rhot_def_with_scattering}], we can apply the eigenvalue decomposition of the Hamiltonian to the unitary to give
\begin{equation}
    U(t, 0) = V_L W^{\dag}e^{iD\tau_L}W V_{L-1}W^{\dag}e^{iD\tau_{L-1}}W\ldots V_2W^{\dag}e^{iD\tau_2}WV_1W^{\dag}e^{iD\tau_1}W.
\end{equation}
Differentiating with respect to $\mathbf{v}$ yields
\begin{equation} \label{eq: partial u U}
    \partial_\mathbf{v} U(t, 0) = \sum_l U(t, t_{l})V_l\bigg(i\tau_lW^{\dag} \partial_\mathbf{v}  D e^{i\tau_l D}W + \partial_\mathbf{v}  W^\dagger e^{-i\tau_l D} W +  W^\dagger e^{-i\tau_l D} \partial_\mathbf{v}  W \bigg) U(t_{l-1},0),
\end{equation}
where $t_l=\sum_{l'\leq l}\tau_{l'}$ is the total duration of time-evolution before the $l^{\text{th}}$ interleaved unitary.
Since multiplication by a unitary operator does not change the spectral norm, we can bound the spectral norm of $A_{\mathbf{v}}$ by the spectral norms of the terms in parentheses.
Applying the triangle inequality  gives
\begin{align}
\|A_{\mathbf{v}}\|_s = \bigg\|\partial_\mathbf{v} U(t, 0)\bigg\|_s\leq \sum_l\Bigg( \tau_l \|\partial_\mathbf{v}  D \|_s + 2 \|\partial_\mathbf{v}  W\|_s \Bigg)= t \|\partial_\mathbf{v}  D\|_s + 2L \|\partial_\mathbf{v}  W\|_s.
\end{align}

We can improve this bound by invoking a second upper bound on $\|A_{\mathbf{v}}\|_s$ and taking the minimum of the two. 
Note that the term in parentheses in Eq.\eqref{eq: partial u U} is equal to
\begin{equation}
    \partial_\mathbf{v}  e^{-iH(\mathbf{u})\tau_l} = -i \int_0^{\tau_l} ds \, e^{-i H(\mathbf{u}) (\tau_l-s)}  \partial_\mathbf{v}  H  e^{-i H(\mathbf{u}) s},
\end{equation}
which is upper bounded by $\| \partial_\mathbf{v}  e^{-iH(\mathbf{u})\tau_l} \|_s \leq \tau_l \| \partial_\mathbf{v}  H(\mathbf{u}) \|_s$.
Applying the triangle inequality to Eq.~(\ref{eq: partial u U}) gives
\begin{equation}
    \|A_{\mathbf{v}}\|_s 
    \leq \| \partial_\mathbf{v}  H(\mathbf{u}) \|_s  \sum_l  \tau_l 
    = t  \| \partial_\mathbf{v}  H(\mathbf{u}) \|_s.
\end{equation}
Inserting the minimum of the two bounds into Eq.~(\ref{eq: I less than A2}) yields the final result.\qed

In the above lemma, the contribution of eigenvalues $D$ to the quantum Fisher information grows quadratically in time, while the contribution of the eigenvectors $W$ eventually saturates to a constant.
Intuitively, this tells us that eigenvalues of the Hamiltonian can be learned at the Heisenberg limit, while eigenvectors cannot.
We formalize this in the following theorem.

\begin{theorem}\label{thm:QFI_bound_unitary_derivative}
Consider the Hamiltonian learning problem (Definition~\ref{def:learning_problem_formal}) for an unknown Hamiltonian $H(\mathbf{u})$ and desired error $\epsilon$, and write the eigenvalue decomposition of the Hamiltonian as $H(\mathbf{u})=W(\mathbf{u})^{\dag} D(\mathbf{u}) W(\mathbf{u})$.
Suppose that each experiment involves discrete quantum control with up to $L$ interleaved unitaries per experiment, and that there exists a normalized vector $\mathbf{v} \in \mathbb{R}^{N_p}$ such that $\partial_{\mathbf{v}} D = 0$.
Then learning the linear combination of parameters $\mathbf{u} \cdot \mathbf{v}$ to within RMS error $\epsilon$ requires a total time
\begin{equation}\label{eq:bound_unitary_invariance}
    T \geq \Omega \left( \epsilon^{\frac{-1}{1+\alpha}} \right), \,\,\,\,\,\,\, \text{if} \,\,\,\,\,\,\, L = \mathcal{O}(T^\alpha).
\end{equation}
In particular, the combination cannot be learned at the Heisenberg limit unless $L = \Omega(T)$, i.e.~$\alpha \geq 1$.
\end{theorem}
\noindent Note that the bound in Eq.~(\ref{eq:bound_unitary_invariance}) is superseded by the Heisenberg limit whenever $\alpha > 1$.
Theorem~2 of the main text follows from this result by setting $L = \mathcal{O}(1)$, i.e.~$\alpha=0$, for which we have $T 
 = \Omega(\epsilon^{-2})$.

\emph{Proof of Theorem~\ref{thm:QFI_bound_unitary_derivative}---}For $\mathbf{v}$ satisfying the conditions of the theorem, application of Lemma~\ref{lem:Hamiltonian_derivative_bound} to each experiment $x$ yields
\begin{equation}
\mathbf{v}^T \bs{\mathcal{I}}^{(Q)} \mathbf{v} \leq \sum_x \min \bigg( 2L\|\partial_{\mathbf{v}} W\|_s\,\, , \,\, t_x \|\partial_{\mathbf{v}}H\|_s\bigg)^2.\label{eq:lambda_min_bound}
\end{equation}
For fixed $T=\sum_x t_x$, this quantity is maximized by choosing $t_x \leq 2 L\|\partial_{\mathbf{v}}W\|_s /\|\partial_{\mathbf{v}}H\|_s$
across $\lceil T/t_x \rceil$ experiments.
Setting $T$ to a multiple of $t_x$ for convenience, this yields
\begin{align}
    \mathbf{v}^T \bs{\mathcal{I}}^{(Q)} \mathbf{v} \leq 4 L^2 \|\partial_{\mathbf{v}} W\|_s^2 \left( \frac{T \| \partial_\mathbf{v} H \|_s}{2 L \|\partial_{\mathbf{v}}W\|_s} \right) = 2 L \|\partial_{\mathbf{v}} W\|_s \|\partial_{\mathbf{v}} H\|_s T.
    \label{eq:lambda_min_bound_optimised}
\end{align}
Following the Cramer-Rao bound, Eq.~(\ref{eq:our_CR_bound}), we have
\begin{equation}
\epsilon^{2} \geq \mathbf{v}^T[\bs{\mathcal{I}}^{(Q)}]^{-1} \mathbf{v} \geq  ( \mathbf{v}^T\bs{\mathcal{I}}^{(Q)} \mathbf{v} )^{-1},
\end{equation}
where the latter bound follows from the Cauchy-Schwarz inequality, as
\begin{equation}
    1 = \mathbf{v}^T [\bs{\mathcal{I}}^{(Q)}]^{1/2}[\bs{\mathcal{I}}^{(Q)}]^{-1/2} \mathbf{v} \leq  \lVert [\bs{\mathcal{I}}^{(Q)}]^{1/2} \mathbf{v} \rVert_2\lVert [\bs{\mathcal{I}}^{(Q)}]^{-1/2} \mathbf{v} \rVert_2 = \sqrt{(\mathbf{v}^T\bs{\mathcal{I}}^{(Q)}\mathbf{v})(\mathbf{v}^T[\bs{\mathcal{I}}^{(Q)}]^{-1}\mathbf{v})}.
\end{equation}
Combining with Eq.~\eqref{eq:lambda_min_bound_optimised}, we have
\begin{equation}
    \epsilon^{2} \geq (2 L \|\partial_{\mathbf{v}} W\|_s \|\partial_{\mathbf{v}} H\|_s T)^{-1}.
\end{equation}
Finally, we can re-arrange terms and set $L = \mathcal{O}(T^\alpha)$ to find
\begin{equation}
        T \geq \Big( 2 \|\partial_{\mathbf{v}}W\|_s\|\partial_{\mathbf{v}}H\|_s \epsilon^{2}\Big)^{\frac{-1}{1+\alpha}},
\end{equation}
as desired. \qed

We conclude by outlining a more physical picture for where the above theorem applies (as was discussed briefly in the main text).
Namely, suppose, as above, that the eigenvalues $D$ of a Hamiltonian $H$ are invariant under an infinitesimal parameter shift, $\mathbf{u}\rightarrow\mathbf{u}+\delta\mathbf{u}$, for some $\delta \mathbf{u}$. This implies the two Hamiltonians are unitarily equivalent: $H(\mathbf{u}+\delta\mathbf{u})=U^{\dag}(\delta \mathbf{u})H(\mathbf{u})U(\delta \mathbf{u})$.
This is a one-to-one relationship, since unitary equivalence in turn implies that $\partial_{\delta \mathbf{u}} D = 0$.
We can therefore state the above result in a more intuitive way: if we can unitarily transform between two nearby Hamiltonians, then we cannot distinguish between them at the Heisenberg limit.
To formalize this, consider a orbit in parameter space along which the spectrum of a Hamiltonian remains invariant.
The presence of such a orbit will preclude Heisenberg-limited learning of Hamiltonians lying along the orbit.

We mention two families of Hamiltonians that feature unitary equivalences---single-qubit Hamiltonians and interacting fermion Hamiltonians---in the main text.
Here, we mention that unitary equivalence also holds for interacting spin Hamiltonians of the form:
\begin{equation}
    H(\mathbf{u})=\sum_{i}u^{(1)}_{i}Z_i + \sum_{\langle i,j\rangle}\Big(u^{(2)}_{ij,xx} X_iX_j + u^{(2)}_{ij,xy}X_iY_j + u^{(2)}_{ij,yx}Y_iX_j + u^{(2)}_{ij,yy}Y_iY_j\Big),
\end{equation}
where $\langle\cdot,\cdot\rangle$ denotes nearest neighbour pairs on a grid.
This possess orbits at any $\mathbf{u}$ that are unitarily-equivalent up to local $Z$-rotations.
This can be extended to arbitrary local rotations by adding other 1- and 2-qubit terms on nearest neighbours.

\section{Bounds on learning thermalizing Hamiltonians} \label{app:ETH_no_learning}

In this section we formally state and prove Theorem~3 of the main text, establishing that Heisenberg-limited Hamiltonian learning is not possible without quantum control in systems that obey the eigenstate thermalization hypothesis (ETH).
We remark that, even on physical grounds, the existence of this result is not clear a priori.
On the one hand, for typical initial states, time-evolution under a thermalizing Hamiltonian will cause the system to approach an equilibrium state where few-body observables are  well approximated by a thermal density matrix.
Since the equilibrium state is independent of time, this appears to preclude learning at the Heisenberg limit.
On the other hand, all Hamiltonian systems can support long-lived coherences within the energy eigenbasis.
The dependence of these coherences on Hamiltonian parameters could in principle open the door to Heisenberg-limited learning.

In what follows, we resolve these questions using the ETH.
We begin in Appendix~\ref{app:ETH} with a formal statement of the ETH, adapted from the standard but non-rigorous definitions in Refs.~\cite{srednicki1994chaos,rigol2008thermalization,d2016quantum,deutsch2018eigenstate}.
In Appendix~\ref{sec:low_rank}, we provide several preliminary definitions and results regarding low-rank approximations of the Fisher information matrix $\bs{\mathcal{I}}^{(Q)}$, as well as the operators $A_a$ defined in Eq.~\eqref{eq:time_integrated_perturbation_def}, which will be useful later on.
In particular, we prove that the existence of a sufficiently low-rank approximation forbids learning all $N_p = \Omega(N)$ Hamiltonian parameters at the Heisenberg limit.
This motivates our main result in Appendix~\ref{app:ETH_formal_proof}: a proof of Theorem~\ref{thm:multi_parameter_ETH_no_HL}, the formal version of Theorem~3 in the main text.
Our proof strategy is to leverage the connection between the quantum Fisher information matrix and the operators $A_a$ defined in Eq.~\eqref{eq:time_integrated_perturbation_def}.
The operators $A_a$ are equal to integrals over time-evolved local Hamiltonian terms, which can be treated via the ETH.

Our proof of Theorem~\ref{thm:multi_parameter_ETH_no_HL} builds upon two lemmas, which we prove separately in Appendixs~\ref{app:ETH_off_diag_bound} and~\ref{app:ETH_diagonal_Aj_rank}.
In Appendix~\ref{app:ETH_off_diag_bound}, we show that, if a Hamiltonian obeys the ETH, the spectral norm of the off-diagonal components of $A_a$ are upper bounded by $\mathcal{O}(t^{1/2})$ and thus cannot contribute to learning at the Heisenberg limit.
Our proof utilizes standard techniques from random matrix theory.
In Appendix~\ref{app:ETH_diagonal_Aj_rank}, we show that the diagonal components of $A_a$ are highly linearly dependent for different parameters $a$, which we capture formally using the notion of low-rank approximations introduced in Appendix~\ref{sec:low_rank}.
This result relies on a physical assumption, namely that the temperature derivatives of thermal expectation values are bounded by a constant (i.e.~do not diverge in the system size).
In Appendix~\ref{app: bounded derivative}, we prove that this assumption is satisfied in any geometrically local Hamiltonian with exponential decay of correlations for operators that enter the Hamiltonian.
Exponential decay of correlations have in turn been established for one-dimensional Hamiltonians at any finite temperature~\cite{bluhm2022exponential} and higher-dimensional Hamiltonians at sufficiently large temperatures~\cite{kliesch2014locality}.
We expect our assumption to hold more broadly than is currently provable (e.g. in all-to-all interacting systems as well), as long as the states of interest are a non-zero temperature difference away from any finite temperature phase transitions.

The above results concern learning many-body Hamiltonians with $N_p = \Omega(N)$ unknown parameters.
In Appendix~\ref{app:ETH_single_parameters}, we conclude by discussing learning in thermalizing Hamiltonians with only a constant number, $N_p = \mathcal{O}(1)$, of unknown parameters.
In this case, we show that Heisenberg-limited Hamiltonian learning may in fact be possible even without quantum control, but requires the ability to prepare and measure coherences between states with extensive differences in energy.
Relatedly, we show that Heisenberg-limited learning is not possible if the initial states are related to product states by any finite depth quantum circuit.

\subsection{The eigenstate thermalization hypothesis} \label{app:ETH}

We begin by formulating a precise statement of the eigenstate thermalization hypothesis.
Thermalization is the common behaviour of observables in physical systems to relax towards a thermal average, regardless of their initial state.
This is surprisingly difficult to predict in quantum mechanics, as the absolute values of a density matrix in the eigenbasis of a system's Hamiltonian $H$ are invariant in time.
A popular method to explain this has been the eigenstate thermalization hypothesis~\cite{srednicki1994chaos,rigol2008thermalization,d2016quantum,deutsch2018eigenstate}.

The eigenstate thermalization hypothesis is commonly stated as follows~\cite{d2016quantum}.
Consider a Hamiltonian $H$ with energy eigenstates $H|E_j\rangle=E_j|E_j\rangle$, and a $k$-local operator $V$ with $k = \mathcal{O}(1)$.
We are interested in the matrix elements of $V$ in the energy eigenbasis.
The eigenstate thermalization hypothesis posits that:
\begin{enumerate}[label=(\roman*)]
\item The diagonal matrix elements of $V$ are given by their thermal expectation values\footnote{To be maximally precise, the diagonal matrix elements are expected to display small random variations $r^V_{ii} \sim \mathcal{O}(e^{-S/2})$ about their thermal values. However, these contributions have exponentially small spectral norm and are thus negligible for our analysis. Nevertheless, if desired, they are easily incorporated into our proof of Theorem~\ref{thm:multi_parameter_ETH_no_HL} by modifying the operators $A_{\text{od},a}$ in Eq.~(\ref{eq:Aod}) to include the diagonal fluctuations.}
\begin{equation} \label{eq:ETH_diagonal}
    \langle E_j|V|E_j\rangle= \expval{V}_{\beta(E_j)},
\end{equation}
where we define an inverse temperature, $\beta(E)$, for each energy via $\expval{H}_{\beta(E)} = E$, where $\expval{\cdot}_{\beta}$ denotes the trace with the thermal density matrix, $\rho_\beta = e^{-\beta H}/\mathrm{Trace}(e^{-\beta H})$. (Such a temperature exists for every energy $E$ in the spectrum of $H$.)

\item The off-diagonal matrix elements of $V$ are given by
\begin{equation}\label{eq:ETH_off_diagonal}
    \langle E_i|V|E_j\rangle = e^{-S(E_{ij})/2}f_V\Big(E_{ij},\omega_{ij}\Big) \, r^V_{ij}, \,\,\,\,\,\,\,\, i \neq j
\end{equation}
where $E_{ij} \equiv \frac{1}{2}(E_i+E_j)$ is the average energy, $\omega_{ij} \equiv E_i-E_j$ is the energy difference, $r^V_{ij} = (r_{ji}^V)^*$ are random\footnote{As has been previously noted~\cite{chen2021fast,dymarsky2022bound}, it is not immediately obvious how to interpret the randomness in $r^V_{ij}$, since the actual matrix elements of interest are defined for a specific, non-random Hamiltonian.
An intuitive way to reconcile this difference is to imagine adding small, random perturbations to the Hamiltonian.
If these perturbations are large compared to the level spacing but small compared to the time-scales of interest, then they can serve to randomize the matrix elements while leaving physical observables approximately unchanged.
Since the level spacing is exponentially small in the system size, there is an exponentially large time window in which this picture can apply.} complex numbers with zero mean and unit variance, and $S(E)$ and $f_V(E,\omega)$ are smooth functions defined below. 
\end{enumerate}
Statement (i) implies that  eigenstates ``look'' thermal from the perspective of observables $V$.
Numerical experiments have found that this indeed holds, up to corrections that are exponentially small in the system size (see~\cite{d2016quantum} for a comprehensive review of numerical experiments).
Statement (ii) implies that the off-diagonal matrix elements ``look'' like random complex numbers, up to the requirement that they reproduce the auto-correlation function of the operator $V$.
Here, $e^{S(E)}$ is the density of states at energy $E$ (which is expected to be exponential in the system size at any finite temperature).
A straightforward calculation shows that the function $f_V(E,\omega)$ is related to the Fourier transform of the connected auto-correlation function $G^c(t;E_i)$ in an eigenstate $\ket{E_i}$ via~\cite{d2016quantum}
\begin{equation}
    |f_V(E_{ij},\omega_{ij})|^2 = e^{S(E_j) - S(E_{ij})} \int_{-\infty}^{\infty} dt \, e^{-i\omega_{ij} t} G^c(t;E_i),
\end{equation}
where we define
\begin{equation} \label{eq: Gc E}
    G^c(t;E_i)= \bra{E_i} V(t)V(0) \ket{E_i} - \bra{E_i} V \ket{E_i}^2,
\end{equation}
and $V(t) = U(t,0)^\dagger V U(t,0)$.
In previous literature it is common to equate the above auto-correlation function with its value in a thermal state [i.e.~replacing $\bra{E} (\cdot) \ket{E} \rightarrow \text{Trace}( (\cdot) \rho_{\beta(E)})$], as well as to approximate the difference in entropies using the inverse temperature  via $S(E_j) - S(E_{ij}) \approx \beta(E_j) \omega_{ij} / 2$.
However, we will not need these simplifications in this work.
Numerical studies have found that the magnitudes of off-diagonal matrix elements are indeed well-described by Eq.~(\ref{eq:ETH_off_diagonal}) (see again~\cite{d2016quantum} for a review).
However, several more recent works have pointed out that the coefficients $r^V_{ij}$ are not in general i.i.d. random numbers; indeed, correlations between different coefficients are in fact necessary to capture the behavior of higher-point correlation functions in the theory~\cite{foini2019eigenstate,chan2019eigenstate,murthy2019bounds,dymarsky2022bound,wang2022eigenstate,jafferis2022matrix}.
After a time-scale $T_{\text{rmt}} \sim \text{poly}(N)$ that is polynomial in the system size, these correlation functions are expected to fully decay and the coefficients to be treatable as i.i.d. random numbers~\cite{cotler2017chaos,schiulaz2019thouless,cotler2020spectral,dymarsky2022bound}.

Motivated by these observations, in this work we adopt the following definition of ETH.
\begin{definition}\label{def:ETH}
    Consider a Hamiltonian $H$ acting on $N$ qubits, with energy eigenstates $H|E_j\rangle=E_j|E_j\rangle$.
    We say that $H$ satisfies the eigenstate thermalization hypothesis (ETH) over an inverse temperature range $[\beta_1,\beta_2]$, if the following conditions hold for all energy eigenvalues $E_i, E_j$ such that $\beta(E_i),\beta(E_j) \in [\beta_1,\beta_2]$:
    \begin{enumerate}
    \item For Pauli operators $V$ that act on a constant number of qubits, time-integrated operators $V(t)$ are well-approximated by random matrix theory predictions,
    \begin{equation}\label{eq:def_ETH}
        \left( \frac{1}{T} \int_0^T dt \, {V}(t) \right) = \left( \frac{1}{T} \int_0^T dt \, {V}_{\text{\textnormal{rmt}}}(t) \right) + \delta {V}, \,\,\,\,\,\,\, \text{with} \,\,\,\,\,\,\,\, \lVert \delta {V} \rVert_s \leq \frac{C}{T},
    \end{equation}
    where ${V}_{\text{\textnormal{rmt}}}$ is defined via Eq.~(\ref{eq:ETH_diagonal}) and Eq.~(\ref{eq:ETH_off_diagonal}) by taking $r^V_{ij}$ as i.i.d. normal random numbers with unit variance.
    Here, $C = \mathcal{O}(\text{poly}(N))$ is a constant independent of $T$.

    \item The system thermalizes in the conventional sense: namely, auto-correlation functions decay to their thermal values after a time-scale that is polynomial in the system size. We capture this via the following condition:
    \begin{equation} \label{eq: Gc E cond}
        \frac{1}{T} \int_0^T dt \, G^c(t;E_i) \leq \frac{C}{T},
    \end{equation}
    for every $\ket{E_j}$.
    We take the constant $C = \mathcal{O}(\text{poly}(N))$ to be equal to that in Eq.~(\ref{eq:def_ETH}) for convenience.

    \item The density of states is exponential in the system size, $e^{S(E)} = \Omega(\text{exp}(N))$.
    \end{enumerate}
\end{definition}

\noindent In conditions both 1 and 2, we find it more natural to state the requirements for thermalization in terms of averages over time-evolved operators.
In condition 1, we note that the time-average leads to a \emph{weaker} condition than the conventional formulation of ETH in Eqs.~(\ref{eq:ETH_diagonal},~\ref{eq:ETH_off_diagonal}), which involves the operator without time-averaging.
This modification enforces that any correlation function involving the approximated operator will be dominated by time differences of order $T$, where we expect the random matrix prescription to be appropriate for sufficiently large times\footnote{Several previous works~\cite{chen2021fast} have addressed this issue in an alternate fashion, by only assuming that ETH holds within a narrow energy band $\omega_{ij} \leq 1/T_{\text{rmt}}$.
Our definition has the benefit that it does not involve a sharp energy cut-off (which, when Fourier transformed, would involve integrating over times ranging up to the inverse level spacing).} (i.e.~$T \gtrsim T_{\text{rmt}}$; see the discussion above Definition~\ref{def:ETH}).
We expect the corrections, $\delta {V}$, to this approximation to scale as at most, $\lVert V \rVert_s \lesssim T_{\text{rmt}}/T$.
We therefore expect the constant $C$ in the definition above to be proportional to the random matrix time scale, $T_{\text{rmt}}$.
In condition 2, we expect the the connected auto-correlation function itself, $G^c(t;E_i)$, to be zero for nearly all times after the ``thermalization time''~\cite{d2016quantum,cotler2022fluctuations}, which occurs before the random matrix time scale $T_{\text{rmt}}$.
The average over time is useful to tolerate Poincare recurrences in the correlation function, which are expected to be exponentially rare in the system size.

Several recent extensions of the ETH go beyond the systems and time regimes captured by the above definition.
Firstly, the above definition fails to describe thermalizing Hamiltonians that possess a conserved quantity besides the energy (for instance, a conserved particle number or a conserved spin parity).
When the additional conserved quantities are mutually commuting, they can be accounted for straightforwardly by applying the ETH to each charge or parity sector independently; additional subtleties are encountered when the conserved quantities are non-commuting~\cite{murthy2022non}.
We expect on physical grounds that our results in the following sections will carry over to both cases, but we do not explore this here.
Secondly, recent work has made tremendous progress towards capturing the correlations between matrix elements at times before $T_{\text{rmt}}$ using non-Gaussian random matrix ensembles~\cite{jafferis2022matrix}.
It would interesting and non-trivial to explore whether our results can be extended using this framework.

\subsection{Low-rank matrix approximations}\label{sec:low_rank}

Similar to Appendix~\ref{app:no_control_no_heisenberg}, our eventual strategy for bounding learning in thermalizing Hamiltonians with no quantum control will utilize the quantum Fisher information matrix, $\bs{\mathcal{I}}^{(Q)}$ [Eq.~\eqref{eq:QFI_pure_states}].
Recall that in order for all $N_p$ parameters of the Hamiltonian to be learnable at the Heisenberg limit, all $N_p$ eigenvalues of the Fisher information matrix must scale as $\Omega(T^2)$.
In the subsequent sections, we will show that for Hamiltonians that obey the ETH nearly all eigenvalues of $\bs{\mathcal{I}}^{(Q)}$ are instead $\mathcal{O}(T)$, and only a sub-extensive number of the eigenvalues scale as $\Omega(T^2)$.

In this section, we set up a few preliminary definitions to capture this behavior.
To maintain generality across different experimental initial states and discrete control operations, we will find it convenient to work with the vector of operators $\bs{A} = (A_1,\ldots,A_{N_p})$ defined in Eq.~(\ref{eq:time_integrated_perturbation_def}) instead of the quantum Fisher information matrix itself.
For an initial state $\rho$, the vector $\bs{A}$ upper bounds the Fisher information matrix via
\begin{equation}\label{eq:QFI_no_second_term reprint}
    \bs{\mathcal{I}}^{(Q)} \leq 4\langle\bs{A} \bs{A}^T\rangle_{\rho},
\end{equation}
which corresponds to dropping the second term in Eq.~(\ref{I A definition vector}), since it is negative semidefinite.
In thermalizing Hamiltonians, the aforementioned sub-extensive number of large Fisher information matrix eigenvalues will be inherited from linear dependencies among the operators $A_a$.
We can see how this might occur with a simple example. 
Suppose that the operators $A_{a}$ are identical for every parameter $a$ up to a multiplicative constant, i.e.~$A_{a}=b_a A$ for all $a$, or in vector form, $\bs{A} = \bs{b}A$.
From Eq.~\eqref{eq:QFI_no_second_term reprint} we have $\bs{\mathcal{I}}^{(Q)} \leq  4 \langle A^2\rangle_\rho \, \bs{b}\bs{b}^T$, which is a rank-$1$ matrix with a single eigenvector, $\bs{b}$.
While this example represents an extreme case, we will find that in thermalizing Hamiltonians, approximate linear dependencies among the $A_a$ significantly restrict learning regardless of the initial state or discrete control operations.

We formalize the linear dependencies above via the notion of a `low-rank' approximation of the vector of operators~$\bs{A}$:
\begin{definition}\label{def: operator rank}
Consider a vector of $N_p$ operators, ${\boldsymbol{A}} = ({A}_1,\ldots, {A}_{N_p})$, and a set of $R$ linearly independent vectors of complex numbers, $\{ \bs{b}_\alpha \in \mathbb{C}^{N_p} \}_{\alpha = 1, ..., R}$. 
We say that $\{ \bs{b}_\alpha \}$ form a rank-$R$ approximation of ${\boldsymbol{A}}$ with error $\delta$, if $\bs{A}$ can be decomposed as follows:
\begin{equation} \label{eq: delta rank}
    {\bs{A}} = \sum_{\alpha=1}^R \bs{b}_\alpha {B}_\alpha + {\bs{E}}, \,\,\,\,\,\,\,\, \text{i.e.} \,\,\,\,\,\,\,\, {A}_a = \sum_{\alpha=1}^R (b_\alpha)_a {B}_\alpha + {E_a},
\end{equation}
where $\{ B_\alpha \}$ are a set of linearly independent operators, and $\bs{E}$ is a vector of operators that obeys
\begin{equation} \label{eq: error bound low rank approximation}
    \lVert \boldsymbol{v}^T {\boldsymbol{E}} \rVert_s = \left\lVert \sum_a v_a {E}_a \right\rVert_s \leq \delta,
\end{equation}
for any normalized vector $\boldsymbol{v} \in \mathbb{R}^{N_p}$, $\boldsymbol{v}^T \boldsymbol{v} = 1$. We say that the decomposition is orthonormal if the $\bs{b}_{\alpha}$ vectors are orthonormal.
\end{definition}
\noindent Intuitively, the low-rank approximation captures the linearly dependencies among $A_1,\ldots,A_{N_p}$ up to deviations less than $\delta$. 
From Eq.~(\ref{eq:QFI_no_second_term reprint}), the existence of a rank-$R$ approximation of $\bs{A}$ immediately implies that the associated quantum Fisher information matrix has at most $R$ eigenvalues greater than $4\delta^2$.

Note that we can apply the Gram-Schmidt procedure to orthonormalize the vectors $\bs{b}_{\alpha}$, by performing an appropriate linear transformation on the operators $B_\alpha$.
We can also assume that the error term, $\bs{E}$, is orthogonal to the vectors $\bs{b}_{\alpha}$, i.e.~$\bs{b}_\alpha^T \bs{E}' = 0$, without loss of generality.
This follows from the following lemma:
\begin{lemma}\label{lem:orthogonalized_low_rank_decompositions}
Consider a set of vectors $\bs{A}$ with a rank-$R$ orthonormal decomposition
\begin{equation}
    \bs{A}=\sum_\alpha\bs{b}_{\alpha}B_{\alpha} + \bs{E},
\end{equation}
with error $\delta$.
The decomposition
\begin{equation}\label{eq:orthogonalizing_decomposition}
    \bs{A}=\sum_\alpha\bs{b}_{\alpha}B'_{\alpha} + \bs{E}',\hspace{1cm} B'_{\alpha}=B_{\alpha}+\bs{b}_{\alpha}^T\bs{E},\hspace{1cm}\bs{E}'=\bs{E}-\sum_{\alpha}\bs{b}_{\alpha}\bs{b}_{\alpha}^T\bs{E},
\end{equation}
is also a rank-$R$ orthonormal decomposition of $\{\bs{A}\}$ with error $\delta$.
Moreover, the latter decomposition obeys $\bs{b}_\alpha^T \bs{E}' = 0$ for all $\alpha$.
\end{lemma}
\emph{Proof}---One can immediately verify that Eq.~\eqref{eq:orthogonalizing_decomposition} is a rank-$R$ decomposition of $\{\bs{A}\}$, and that $\bs{b}^T_\alpha \bs{E}' = 0$.
It remains to show that the adjustment $\bs{E} \rightarrow \bs{E}'$ does not increase the approximation error $\delta$.
To show this, we decompose an arbitrary vector $\bs{v}$ as $\bs{v} = \bs{v}_\parallel + \bs{v}_\perp$, where $\bs{v}_\parallel \in \text{span} \{ \bs{b}_{\alpha} \}$ and $\bs{v}_\perp$ is orthogonal to $\text{span} \{ \bs{b}_{\alpha} \}$.
Since $\bs{b}^T_\alpha \bs{E}' = 0$, we have $\bs{v}^T \bs{E}' = \bs{v}_\perp^T \bs{E}' = \bs{v}_\perp^T \bs{E}$.
Taking the infinity norm gives the desired approximation error bound, $\lVert \bs{v}^T \bs{E}' \rVert_s = \lVert \bs{v_\perp}^T \bs{E} \rVert_s \leq \delta$.\qed

Let us now extend the this framework to a learning protocol involving multiple experiments $x$, associated vectors of operators, $\bs{A}_x$, and a total quantum Fisher information matrix, $\bs{\mathcal{I}}^{(Q)} = \sum_x \bs{\mathcal{I}}^{(Q)}_x$ (see Appendix~\ref{app:background} for definitions of each quantity).
Now, suppose that the vectors $\bs{A}_x$ in each experiment have a low-rank decomposition, and moreover, that the vectors $\bs{b}_\alpha$ are \emph{shared} between the decompositions for each experiment.
In this case, Eq.~(\ref{eq:QFI_no_second_term reprint}) again immediately implies that the total quantum Fisher information matrix has at most $R$ eigenvalues greater than $4 \sum_x \delta_x^2$, where the errors $\delta_x$ in each approximation may add in quadrature.
This holds even if the operators $B_{\alpha,x}$ and error terms $\bs{E}_{x}$ vary from experiment to experiment.

With this intuition, let us extend the definition of a low-rank approximation from a single vector $\bs{A}$ to a set of vectors $\{\bs{A}_x\}$ corresponding to different experiments $x$.
\begin{definition}\label{def:low_rank_decomposition}
Consider a set of vectors of $N_p$ operators, $\{ \bs{A}_x \}$, and a set of $R$ linearly independent vectors of complex numbers, $\{ \bs{b}_\alpha \in \mathbb{C}^{N_p} \}_{\alpha = 1, ..., R}$.
We say that $\{ \bs{b}_\alpha \}$ form a rank-$R$ approximation of $\{\bs{A}_x\}$ with errors $\{\delta_x\}$, if $\{\bs{b}_{\alpha}\}$ is a rank-$R$ decomposition (Definition~\ref{eq: delta rank}) of $\bs{A}_x$ with error $\delta_x$ for each $x$. That is, there exists operators $B_{x,\alpha}$ and vectors of operators $\bs{E}_x$ such that
\begin{equation}
    \bs{A}_x=\sum_\alpha\bs{b}_{\alpha}B_{x,\alpha} + \bs{E}_x,
\end{equation}
with
\begin{equation} \label{eq: error bound low rank approximation many vectors}
    \lVert \boldsymbol{v}^T {\boldsymbol{E}_x} \rVert_s = \left\lVert \sum_a v_a {E}_{x,a} \right\rVert_s \leq \delta_x,
\end{equation}
for any normalized vector $\boldsymbol{v} \in \mathbb{R}^{N_p}$, $\boldsymbol{v}^T \boldsymbol{v} = 1$.
\end{definition}
\noindent The preceding arguments now lead us to the following lemma:
\begin{lemma} \label{rank lower bounds du}
Consider a set of quantum experiments $x$ with unitary evolutions $U_x$, and define $\bs{A}_x$ for each experiment as in Eq.~(\ref{I A definition vector}).
Suppose that $\{\bs{b}_\alpha\}$ forms a rank-$R$ approximation for $\{\bs{A}_x\}$ with errors $\{\delta_x\}$.
Then the total RMS error of any resulting estimate of the parameters is lower bounded by
\begin{equation} \label{eq:error from low rank multiple expts}
     \epsilon \geq \frac{1}{2\delta} \sqrt{1 - R/N_p},
\end{equation}
with $\delta^2 \equiv \sum_x\delta_x^2$.
\end{lemma}

\emph{Proof}---Inserting the low-rank decomposition of $\bs{A}$ into the bound in Eq.~(\ref{eq:QFI_no_second_term reprint}) and invoking Lemma~\ref{lem:orthogonalized_low_rank_decompositions} to fix $\bs{b}^T_{\alpha}\bs{E}_x=0$, we have
\begin{align}
    \bs{\mathcal{I}}^{(Q)} \leq 4 \sum_{x=1}^{N_x}  \expval{ {\boldsymbol{A}}_{x} {\boldsymbol{A}}_{x}^T}_{\Psi_x} &\leq 4 \sum_{x=1}^{N_x} \left( \sum_{\alpha,\beta} \bs{b}_{\alpha} \bs{b}_{\beta}^T \expval{ {B}_{x,\alpha} {B}_{x,\beta}}_{\Psi_x} + \expval{ {\boldsymbol{E}}_x  {\boldsymbol{E}}_x^T }_{\Psi_x} \right)\nonumber\\
    \label{eq: rank lower bounds du}
    &=4 \sum_{\alpha,\beta} \bs{b}_{\alpha} \bs{b}_{\beta}^T \sum_{x=1}^{N_x}\expval{ {B}_{x,\alpha} {B}_{x,\beta}}_{\Psi_x} + \sum_{x=1}^{N_x}\expval{ {\boldsymbol{E}}_x  {\boldsymbol{E}}_x^T }_{\Psi_x} .
\end{align}
Note that we are able to invoke Lemma~\ref{lem:orthogonalized_low_rank_decompositions} simultaneously on all $\bs{A}_x$ as the orthogonalization with the error term only adjusts the operators $\bs{E}_x$ and $B_{x,\alpha}$ (which are unique to each $\bs{A}_x$), and not the shared vectors $\bs{b}_{\alpha}$.
The first term has matrix rank of at most $R$, while the latter term obeys $\boldsymbol{v}^T \langle {\boldsymbol{E}}_x  {\boldsymbol{E}}_x^T \rangle_{\Psi_x} \boldsymbol{v} \leq \delta_x^2$ for any $\bs{v} \in \mathbb{R}^{N_p}$.
Summing over $x$, we see that $\bs{\mathcal{I}}^{(Q)}$ has at most $R$ eigenvalues of magnitude greater than $4 \delta^2$, with $\delta^2 = \sum_x \delta_x^2$.
We can therefore bound the trace of $[\bs{\mathcal{I}}^{(Q)}]^{-1}$ below by
\begin{equation}
    \sum_a  \left[\bs{\mathcal{I}}^{(Q)} \right]^{-1}_{aa} \geq  \frac{N_p - R}{4\delta^2},
\end{equation}
which corresponds to taking the $R$ eigenvalues of $\mathcal{I}$ that are greater than $4\delta^2$ to infinity, and the $N_p-R$ remaining eigenvalues to $4\delta^2$.
Applying Eq.~(\ref{eq:our_CR_bound}) then gives Eq.~(\ref{eq:error from low rank multiple expts}). \qed

We conclude this section by turning directly to Hamiltonian learning.
We consider learning a time-independent Hamiltonian with $L$ discrete quantum control operations (Definition~\ref{def:discrete_quantum_control}).
The case of no quantum control (Definition~\ref{def:no_quantum_control}) corresponds to $L=1$.
Differentiating Eq.~\eqref{eq:rhot_def_with_scattering} via the product rule then yields
\begin{equation} \label{eq: WAW}
    \bs{A}_x = -i\sum_l U^{\dag}_x(t_{x,l-1},0) \bs{A}_H(\tau_{x,l}) U_x(t_{x,l-1},0), \,\,\,\,\,\,\,\,\, \bs{A}_{H}(t) = -i\int_0^{t} ds \, e^{iHs}\bs{P}e^{-iHs},
\end{equation}
where $U(t_{x, l-1},0) = \prod_{l'=1}^{l-1} \left( V_{l'} e^{iH\tau_{x,l'}} \right)$, $t_{x,l}=\sum_{l'\leq l}\tau_{x,l}$ is time evolution up till time $t_{x, l-1}$, and $\bs{P} = (P_1,\ldots,P_{N_p})$ is the vector of Pauli operators that enter $H$.
Note that the operator $\bs{A}_H(\tau_{x,l})$ depends on the experiment $x$ only through the times $\tau_{x,l}$.

We can now state the main result of this section:
\begin{lemma} \label{lemma: W HL}
Consider the Hamiltonian learning problem (Definition~\ref{def:learning_problem_formal}), with discrete control of up to $L$ interleaved unitaries.
Suppose that the vectors $\{ \bs{b}_\alpha\}$ form a rank-$R$ approximation of $\bs{A}_H(t)$,
\begin{equation}
    \bs{A}_{H}(t) = \sum_{\alpha}\bs{b}_{\alpha}A_{\alpha}(t) + \bs{E}(t)
\end{equation}
with error $\delta_H = \sqrt{\mathfrak{a} t}$, for all times $t \geq 0$ with $\mathfrak{a}$ constant.
Then, the maximum root-mean-square error is bounded below as
\begin{equation} \label{eq: W HL 1}
    \epsilon \geq \frac{1}{2\sqrt{\mathfrak{a} L T}} \sqrt{1 - R/N_p},
\end{equation}
where $T$ is the total evolution time.
Hence, the parameters $\boldsymbol{u}$ cannot be learned at the Heisenberg limit if $L = o(T)$ and $R = o(N_p)$.
\end{lemma}
\noindent The above assumes that the vectors $\bs{b}_\alpha$ are time-independent, which we will find is indeed the case in the Hamiltonians we consider.

\emph{Proof of Lemma~\ref{lemma: W HL}}---Inserting the low-rank approximation of $\bs{A}_H(\tau)$ into Eq.~(\ref{eq: WAW}) gives a low-rank approximation for $\{\bs{A}_x\}$:
\begin{equation}\label{eq:low_rank_AH_insertion}
    {\boldsymbol{A}}_x = \sum_{l,\alpha} \bs{b}_{\alpha}\left[ U^{\dag}_x(t_{x,l-1},0) {A}_{\alpha}(\tau_{x,l}) U_x(t_{x,l-1},0) \right] + \sum_l \left[ U^{\dag}_x(t_{x,l-1},0) {\boldsymbol{E}}(\tau_{x,l}) U_x(t_{x,l-1},0) \right].
\end{equation}
The latter term has an operator norm of at most $\delta_x = \sum_l \sqrt{\mathfrak{a} \tau_{l,x}} \leq \sqrt{\mathfrak{a} Lt_x}$, where $t_x = \sum_l \tau_{x,l}$ is the total evolution time in measurement round $x$.
Summing these error terms over $x$ then yields $\delta^2 = \sum_x \delta_x^2 \leq \mathfrak{a}L \sum_x t_x = \mathfrak{a} L T$.
Application of Lemma~\ref{rank lower bounds du} then gives Eq.~(\ref{eq: W HL 1}). \qed

\subsection{Formal statement and proof of Theorem~\ref{thm:multi_parameter_ETH_no_HL}}\label{app:ETH_formal_proof}

We are now ready to state a formal version of Theorem~3 in the main text.
\begin{theorem} \label{thm:multi_parameter_ETH_no_HL}
    Consider the learning problem (Definition~\ref{def:learning_problem_formal}) for an $N$-qubit Hamiltonian without continuous quantum control [Eq.~\eqref{eq:H_no_control_formal}] and up to $L$ interleaved unitaries in the discrete quantum control model.
    Suppose that, within an inverse temperature range $[\beta_1,\beta_2]$, the Hamiltonian $H(\bf{u})$: (i) obeys the eigenstate thermalization hypothesis as in Definition~\ref{def:ETH}, and (ii) has thermal expectation values with bounded derivatives with respect to the inverse temperature
    \begin{equation} \label{eq:bounded_deriv}
        \left| \frac{ \partial_\beta^m \expval{P_a}_\beta }{ m! } \right| \leq (m^\gamma B)^m, \,\,\,\,\,\,\,\, \gamma, B = \mathcal{O}(1)
    \end{equation}
    for all operators $P_a$ in the Hamiltonian.
    Then the root-mean-square error in the parameters $\bf{u}$ is lower bounded by
    \begin{equation} \label{ETH rms bound}
        \epsilon \geq \frac{1}{2\sqrt{F^* L T}} \sqrt{1 - \frac{2 B |\beta_2-\beta_1| \log_2 \left( \sqrt{N_p T / F^*} \right)^{\gamma+1}}{N_p}},
    \end{equation}
    for any learning protocol that can only access states within $[\beta_1,\beta_2]$.
    Here, $F^* = \text{\emph{poly}}(N)$ is constant in $t$, and the bound holds whenever the term inside the square root is positive.
    Thus, if $L = o(T)$ and $N_p=\Omega(N^c)$ for some $c > 0$, then the Heisenberg limit cannot be achieved for $T=o \! \left( \exp( N^{2c/(\gamma+1)}) \right)$.
\end{theorem}

\noindent In many-body Hamiltonian learning, the number of unknown parameters is typically at least linear in the system size, $N_p = \Omega(N)$.
In this case, the above theorem precludes Heisenberg-limited learning until times exponential in the system size.
[See Appendix~\ref{app:ETH_single_parameters} for a discussion of learning in thermalizing Hamiltonians with only a constant number of unknown parameters, $N_p = \mathcal{O}(1)$.]
The required bound on thermal derivatives is satisfied by most physical systems away from finite temperature phase transitions (where the constant $B$ diverges)~\cite{altland2010condensed}.
In Appendix~\ref{app: bounded derivative}, we leverage previous results~\cite{kliesch2014locality,bluhm2022exponential} to prove that this bound holds for all local one-dimensional Hamiltonians at any finite temperature, and local higher-dimensional Hamiltonians at sufficiently high temperatures.

We note that our lower bound contains a prefactor $1/\sqrt{F^*} \sim 1/\text{poly}(N)$, and thus becomes weaker as the system size increases.
This prefactor arises from our lenient definition of thermalization: We assume only that connected correlation functions decay to zero after some time polynomial in $N$, and place no restriction on their behavior before this time (see the second condition in Definition~\ref{def:ETH}).
We expect that the prefactor can be improved by placing more stringent assumptions on the early-time behavior.

\emph{Proof of Theorem~\ref{thm:multi_parameter_ETH_no_HL}}---The bulk of the proof is relegated to the following two sections.
Here we show how the results of these sections, Lemmas~\ref{lemma:OD} and~\ref{lemma:D}, lead to Theorem~\ref{thm:multi_parameter_ETH_no_HL}.

We begin by using the eigenstate thermalization hypothesis (Definition~\ref{def:ETH}) to decompose the term $\bs{A}_H(t)$ in Eq.~\eqref{eq: WAW} into a sum of three components,
\begin{equation} \label{eq: A decomp}
    \bs{A}_H(t) = \int_0^t ds \, \bs{P}(s) = \bs{A}_{\text{d}}(t) + \bs{A}_{\text{od}}(t) + \delta \bs{A}(t),
\end{equation}
where $\bs{P}$ is the vector of $N_p$ Pauli operators that appear in the Hamiltonian.
For each term above, we set all matrix elements outside of the energy range specified by $[\beta_1,\beta_2]$ to zero, which is allowed by assumption.
The first term corresponds to the diagonal component in ETH,
\begin{equation} \label{eq:Ad}
    A_{\text{d},a}(t) = t \sum_i \expval{P_a}_{\beta(E_i)} \dyad{E_i}.
\end{equation}
The second term corresponds to the off-diagonal component in ETH,
\begin{equation} \label{eq:Aod}
    A_{\text{od},a}(t) = \sum_{i < j} \left( e^{-S(E_{ij})/2} f^{a}(E_{ij},\omega_{ij}) \, \frac{e^{i \omega_{ij} t} - 1}{i \omega_{ij}} \, r^{a}_{ij} \right) \dyad{E_i}{E_j} + h.c.,
\end{equation}
where $r^a_{ij}$ are i.i.d. normal random complex numbers with unit variance.
Here we have performed the time integral, $\int_0^t ds \, e^{i \omega s} = (e^{i \omega t}-1)/(i\omega)$.
Finally, the third term is given by $\delta A_{a}(t) = t \cdot \delta {P}_a(t)$,
where ${P}_a(t)$ quantifies the deviation from ETH behavior in Definition~\ref{def:ETH}.
This term is upper bounded by assumption from Definition~\ref{def:ETH},
\begin{equation}
    \lVert \bs{w}^T \delta \bs{A}(t) \rVert_s \leq C \sqrt{N_p} , \,\,\,\,\,\,\,\, \bs{w}^T \bs{w} = 1,
\end{equation}
where $C = \mathcal{O}(\text{poly}(N))$ is defined via Definition~\ref{def:ETH} and is independent of $t$.
This implies that the third term cannot contribute to parameter learning at the Heisenberg limit.

In Lemma~\ref{lemma:OD} of the following section, we show that the second term can be upper bounded using techniques from random matrix theory,
\begin{equation}
    \lVert \bs{w}^T \bs{A}_{\text{od}}(t) \rVert_s \leq \sqrt{F t}, \,\,\,\,\,\,\,\, \bs{w}^T \bs{w} = 1,
\end{equation}
where $F = \mathcal{O}(\text{poly}(N))$ is a constant independent of $t$ (see Lemma~\ref{lemma:OD} for a definition). 
This implies that the second term cannot contribute to parameter learning at the Heisenberg limit.

In Lemma~\ref{lemma:D} of Appendix~\ref{app:ETH_diagonal_Aj_rank}, we show that the first term, $\bs{A}_{\text{d}}(t)$, possesses a low-rank approximation with error $\delta$ composed of $R(\delta,t) = 2 B | \beta_2 - \beta_1 | \log_2(t\sqrt{N_p}/ \delta)^{\gamma+1}$ \emph{time-independent} vectors\footnote{By `time-independent' we mean e.g. that each vector used in the low-rank approximation of $\bs{A}_d(t)$ appears in the low-rank approximation of $\bs{A}_d(2t)$. However, the low-rank approximation of $\bs{A}_d(2t)$ will contain $\mathcal{O}(1)$ additional vectors than that of $\bs{A}_d(t)$.}.
Combining this fact with the above bounds on the second and third term, we see that the entire operator vector $\bs{A}_H(t)$ possesses the same low-rank approximation with error
\begin{equation}
    \delta^* = \delta + \sqrt{F t} + \sqrt{N_p} C.
\end{equation}
Since the latter two terms are $\mathcal{O}(\sqrt{t})$, there exists a constant $F^*$ such that $\sqrt{F^*t/4} > \sqrt{F t} + \sqrt{N_p} C$ for all $t\geq F$.
We note that $F^* = \text{poly}(N)$ since $F = \text{poly}(N)$ and $N_p = \text{poly}(N)$ (the latter follows because there are at most ${N \choose k} 3^k = \mathcal{O}(N^k)$ possible Pauli terms in a $k$-local Hamiltonian).

The results of the above paragraph demonstrate that $\bs{A}_H(t)$ satisfies the conditions of Lemma~\ref{lemma: W HL}, with error $\delta^* = \sqrt{F^* t}$ and rank $R(\delta^*/2,t)$.
Invoking Lemma~\ref{lemma: W HL} and inserting the definition of $R(\delta,t)$ above yields Eq.~(\ref{ETH rms bound}). 

One can confirm that as long as Eq.~\eqref{ETH rms bound} holds, the Heisenberg limit cannot be achieved unless $L$ scales at least linearly in $T$.
It remains to consider what would be required for this inequality to break down; i.e. for the term within the square root to become negative.
Some quick algebra shows that the inequality holds whenever
\begin{equation}
    \log_2(T) \leq \Bigg[\frac{N_p}{2^{-\gamma}B|\beta_2-\beta_1|}\Bigg]^{\frac{1}{\gamma+1}} + \log_2(F^*) - \log_2(N_p).
\end{equation}
We recall that $F^*$ is constant in $t$ and polynomial in $N$.
Thus, if the number of unknown parameters scales with any non-zero power of the system size, $N_p = \Omega(N^c)$ with $c>0$, then term in the square root is positive up to times exponentially large in a non-zero power of the system size, $T = o(\text{exp}(N^{c/(\gamma+1)}))$. \qed

\subsection{Upper bounding the spectral norm of the off-diagonal component}\label{app:ETH_off_diag_bound}

In this section, we bound the operator norm of the off-diagonal matrix elements via the following Lemma.
\begin{lemma} \label{lemma:OD}
    The vector of operators $\bs{A}_{\text{\textnormal{od}}}$ defined in Eq.~(\ref{eq:def_ETH}) obeys the inequality
    \begin{equation}
        \lVert \bs{w}^T \bs{A}_{\text{\textnormal{od}}} \rVert_s \leq \sqrt{ F t},
    \end{equation}
    for any $\bs{w} \in \mathbb{R}^{N_p}$ with $\bs{w}^T \bs{w} = 1$, with probability double exponentially close to one in the system size.
    Here $F$ is defined via
    \begin{equation}\label{eq:F_def}
        F = 8 \sqrt{2} \, \max_i \int_{-t}^t ds \, \left( 1 -  |s|/t  \right) \cdot G^c(s;E_i),
    \end{equation}
    where $G^c(s;E_i)$ is the connected auto-correlation function [Eq.~(\ref{eq: Gc E})] of the operator $\bs{w}^T \bs{P}$ in the energy eigenstate $\ket{E_i}$ at time $s$.
\end{lemma}
\noindent By condition 2 of Definition~\ref{def:ETH} [Eq.~(\ref{eq: Gc E})], we have $F = \mathcal{O}(\text{poly}(N))$ .

\emph{Proof of Lemma~\ref{lemma:OD}}---Our proof uses several basic results borrowed from random matrix theory~\cite{mingo2017free}.
Let us begin by defining, for a given vector $\bs{w}$, the time-integrated perturbation operator $A \equiv \bs{w}^T \bs{A}_{od}$ and the associated sum of Pauli operators $P \equiv \bs{w}^T \bs{P}$.
The off-diagonal matrix elements of $A$ are normal random complex variables with mean zero and variance,
\begin{equation}
    v_{ij} \equiv \mathbb{E} \left[ |A_{ij}|^2 \right] = e^{-S(E_{ij})} |f(E_{ij},\omega_{ij})|^2 \frac{\sin(\omega_{ij} t/2)^2}{\omega_{ij}^2},
\end{equation}
where $|f|^2 = \sum_a | w_a |^2 | f_a |^2$, and $f_a$ describe the off-diagonal matrix elements of the individual Pauli operator $P_a$.
The matrix $A$ is thus drawn from an ensemble with probability measure $dP$ given by
\begin{equation} \label{eq: dP(A)}
    d P(A) = \left( \prod_{i < j} \frac{dA_{ij}}{\sqrt{2\pi v_{ij}}} \exp \left(- |A_{ij}|^2 / 2 v_{ij} \right) \right).
\end{equation}
Note that the variance depends on the element indices $i,j$.
The Gaussian unitary ensemble typically is defined via $v_{ij} = v/d$, where $v$ is an arbitrary scaling factor and $d$ is the matrix dimension.

\begin{figure*}
    \centering
    \includegraphics[width=\linewidth]{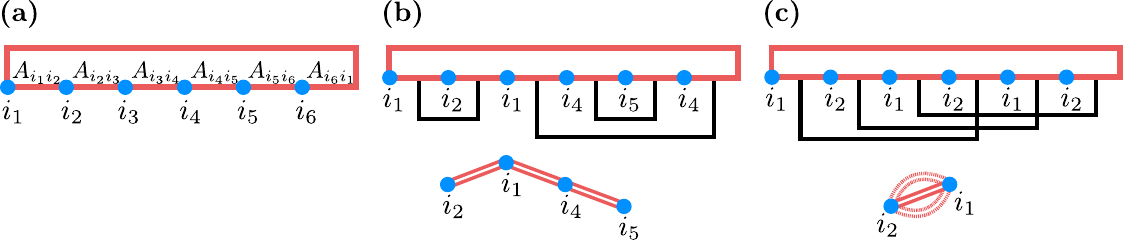}
    \caption{
    \textbf{(a)} Schematic of matrix multiplications and Wick contractions for evaluating $\mu'_{2m} = \text{Trace}(A^{2m})$ with $m=3$.
    Blue dots denote energy indices $i_r$ for $r=1,\ldots,6$ which are summed over.
    Red lines denote matrix elements $A_{i_r i_{r+1}}$.
    \textbf{(b)} An example of a non-crossing Wick contraction.
    Top: The Wick contraction (black lines) pairs matrix elements $A_{i_p i_{p+1}}$ and $A_{i_q i_{q+1}}$ by enforcing the constraints $i_{p+1} = i_q$, $i_{q+1} = i_p$.
    For non-crossing pairings, these constraints leave $m+1$ independent indices to sum over (in this example, indices $i_1,i_2,i_4,i_5$).
    Bottom: The graph $G$ defined by identifying all blue dots with the same index.
    Edges in $G$ are denoted with a double line to signify that they carry a factor of the matrix element squared, $| A_{ij} |^2$.
    Since the contraction is non-crossing, the graph is a tree (see text for further discussion).
    \textbf{(c)} An example of a Wick contraction with crossings, with genus $g = 1$.
    Top: The constraints leave $m+1-2g$ independent indices to sum over (in this example, indices $i_1,i_2$).
    Bottom: The graph $G$ defined by identifying all blue dots with the same index.
    Edges that are not part of the spanning tree $T$ (see text) are denote with striped double lines.
    }
    \label{fig:randommatrixdiagrams}
\end{figure*}

Following established techniques in random matrix theory~\cite{mingo2017free}, the expected moments of $A$,
\begin{equation} \label{mup2m}
    \mu'_{2m} \equiv \mathbb{E} \left[ \mathrm{Trace}( A^{2m} ) \right]
    = \sum_{i_1, \ldots, i_{2m}} \mathbb{E} \left[  A_{i_1 i_2} A_{i_2 i_3}\ldots A_{i_{2m} i_{1}} \right],
\end{equation}
can be calculated as sums over Wick contractions:
\begin{equation}
    \mu'_{2m} = \sum_\pi c_\pi = \sum_\pi \sum_{i_1, \ldots, i_{2m}} \left(  \prod_{(p,q)\in \pi} \mathbb{E} \left[  | A_{i_p i_q} |^2 \right] \cdot  \delta_{i_p,i_{q+1}} \cdot \delta_{i_q,i_{p+1}} \right),
\end{equation}
where the contribution $c_\pi$ of an individual Wick contraction $\pi$ is defined via the second equality.
See Fig.~\ref{fig:randommatrixdiagrams} for a visual representation of the matrix multiplication and Wick contractions.
Each contraction corresponds to a partition $\pi$ of the $2m$ copies of $A$ into $m$ pairs.
Each pair $(p,q) \in \pi$ contributes a factor of $\mathbb{E} \left[  | A_{i_p i_q} |^2 \right] = v_{i_p i_q}$ to the contraction.
Meanwhile, the matrix multiplications and trace enforce the conditions $\delta_{i_p,i_{q+1}}$ and $\delta_{i_q,i_{p+1}}$ for each $p,q$.
Here, the addition should be performed modulo $2m$, to incorporate the condition $\delta_{i_{2m},i_1}$ enforced by the trace. 

In what follows, we will upper bound the sum over all Wick contractions using techniques from random matrix theory.
We refer to~\cite{mingo2017free} for a comprehensive introduction to the relevant random matrix techniques.

We begin by discussing only the Wick contractions that contribute to leading order in the inverse density of states, $e^{-S}$.
The dominant Wick contractions at leading order in $e^{-S}$ correspond to \emph{non-crossing} pairings.
If one orders the $2m$ operators, $A_{i_1 i_2},\ldots,A_{i_{2m} i_1}$, in a circle and draws lines between them for each contraction, a non-crossing pairing contains no lines that cross [Fig.~\ref{fig:randommatrixdiagrams}(b)].
Such a pairing contains `innermost pairs', where $A_{i_{r-1},i_{r}}$ is paired with $A_{i_r,i_{r+1}}$ [for example,  $r = 2,5$ in Fig.~\ref{fig:randommatrixdiagrams}(b)].
An innermost pair in a non-crossing pairing contributes a factor
\begin{equation}
    \sum_{j} \mathbb{E} \left[ | A_{ij} |^2 \right] = \sum_{j} v_{ij},
\end{equation}
to the entire Wick contraction.
This factor can, in fact, be re-expressed in terms of the connected auto-correlation function [Eq.~(\ref{eq: Gc E})] as follows:
\begin{equation} \label{eq:bound G}
\begin{split}
    \sum_{j} v_{ij} & = \int d E_j \, e^{S(E_j)-S(E_{ij})} |f(E_{ij},\omega_{ij})|^2 \cdot \frac{\sin(\omega_{ij} t/2)^2}{\omega_{ij}^2} \\
    & = \sum_j e^{-S(E_{ij})} |f(E_{ij},\omega_{ij})|^2 \cdot \int_0^t dt_1 \int_0^t dt_2 \, e^{i \omega_{ij} (t_1 - t_2)} \\
    & = t \sqrt{2} \int_{-t}^t ds \, \left( 1 -  \frac{|s|}{t}  \right) \cdot G^c(s;\beta(E_i)),
\end{split}
\end{equation}
where the third line follows from re-arranging the integration coordinates and applying the definition of the connected auto-correlation function [Eq.~(\ref{eq: Gc E})].
The sum over $v_{ij}$ is therefore upper bounded by
\begin{equation}
    \sum_{j} v_{ij} \leq \frac{ F \, t}{8}
\end{equation}
from the definition of $F$ [Eq.~(\ref{eq:F_def})].
The product $F t / 8$ serves as the scaling factor of a comparable Gaussian unitary ensemble [see discussion under Eq.~(\ref{eq: dP(A)})].

Using the above lower bound, we can remove the sums over all indices that lie in the center of an innermost pairing from Eq.~(\ref{mup2m}).
This leaves a non-crossing pairing on the remaining indices [for example,  $r = 1,4$ remain after upper bounding innermost pairs in Fig.~\ref{fig:randommatrixdiagrams}(b)].
We can successively iterate the above procedure to bound the total contribution of the original non-crossing pairing by $d (Ft)^m$, where the factor of the Hilbert space dimension, $d$, arises from the final index summation~\cite{mingo2017free}.
Summing over all non-crossing pairings gives the following upper bound on the leading order contribution to the $2m^{\text{th}}$ moment:
\begin{equation} \label{eq: leading order 2m}
\begin{split}
    \left( \mu'_{2m} \right)_{\text{leading order}} & \leq d \, C_m (F t)^m, 
\end{split}
\end{equation}
where $d$ is again the matrix dimension, and $C_m$ is total number of non-crossing pairings of $2m$ elements, also known as the $m^{\text{th}}$ Catalan number~\cite{mingo2017free}.
The right hand side is also equal to the $2m^{\text{th}}$ moment of the Gaussian unitary ensemble with scaling factor $F t /8$.

We now explicitly bound the sum over all Wick contractions.
Consider the contribution of a pairing $\pi$.
The independent energy indices that are summed over are determined by $\pi$ as well as the ordering of the  operators in the trace.
This results in a sum over $k_\pi = m + 1 - 2 g_\pi$ independent energy indices, where the integer $g_\pi$ is known as the \emph{genus} of the pairing $\pi$~\cite{mingo2017free}.
To proceed, let us associate the $k_\pi$ independent energy indices with vertices in a graph $G$, which are connected by edges corresponding to the matrices $A_{i_n,i_{n+1}}$ [see bottom of Fig.~\ref{fig:randommatrixdiagrams}(b,c)].
Note that $G$ is necessarily connected, as $G$ can be viewed as a circular graph with some vertices identified.
The contribution of the pairing $\pi$ can now be written in the form
\begin{equation} \label{eq: cpi}
\begin{split}
    c_\pi & = \left( \prod_{r \in V(G)} \sum_{i_r} \right) \prod_{(p,q) \in E(G)} e^{-S(E_{i_p i_q})} | f_{i_p i_q} |^2 \frac{ \sin( \omega_{i_p i_q} t / 2)^2}{\omega_{i_p i_q}^2} \\
    & = \left( \prod_{r \in V(G)} \int dE_{i_r} \, e^{S(E_{i_r})} \right)  \prod_{(p,q) \in E(G)} e^{-S(E_{i_p i_q})} | f_{i_p i_q} |^2 \frac{ \sin( \omega_{i_p i_q} t / 2)^2}{\omega_{i_p i_q}^2}, \\
\end{split}
\end{equation}
where we denote the vertices and edges of the graph $G$ by $V(G)$ and $E(G)$, respectively.
Each vertex in $G$ contributes a sum over its associated energy index, and each edge in $G$ contributes a factor of $\mathbb{E}[ | A_{i_p i_q} |^2 ]$.

To bound Eq.~(\ref{eq: cpi}),  we consider a spanning tree $T$ of $G$, which will contain $k_\pi$ vertices and $k_\pi - 1$ edges, since each edge in $T$ can be uniquely associated with its child vertex, which leaves only the root vertex without an associated edge.
There are $2g_\pi$ edges that are in $G$ but not in $T$.
Now, for each such edge $(p,q)$, we apply the inequality
\begin{equation}
    e^{- S(E_{i_p i_q})} | f_{i_p i_q} |^2 \frac{\sin(\omega_{i_p i_q} t/2)^2}{\omega_{i_p i_q}^2} \leq \frac{\mathfrak{m} t}{2},
\end{equation}
which is derived by applying the inequality, $\sin^2(x)/x^2 \leq 1$, and defining the maximum average squared off-diagonal matrix element,
\begin{equation}
    \mathfrak{m} = \max_{ij} e^{- S(E_{i j})} | f_{i j} |^2 = \mathcal{O}(\text{exp}(-N)).
\end{equation}
Note that $\mathfrak{m}$ is exponentially small in the system size due to the factor of $e^{-S}$.
This leaves the following expression:
\begin{equation} \label{eq: cpi bound 1}
\begin{split}
    c_\pi & \leq \left( \frac{\mathfrak{m} t}{2} \right)^{2g_\pi} \left( \prod_{r \in V(T)} \int dE_{i_r} \, e^{S(E_{i_r})} \right) \prod_{(p,q) \in E(T)} e^{- S(E_{i_p i_q})} | f_{i_p i_q} |^2 \frac{\sin(\omega_{i_p i_q} t/2)^2}{\omega_{i_p i_q}^2},  \\
\end{split}
\end{equation}
where the RHS differs from Eq.~(\ref{eq: cpi}) because the product is now over edges in $T$ and not $G$.
We recall $V(T) = V(G)$ since $T$ is spanning.

The benefit of the above expression is that every edge term is uniquely associated with a vertex integral because $T$ is a tree.
Namely, each edge can be associated with its child vertex.
Let us denote $p$ as the parent vertex and $q$ the child vertex for each edge $(p,q) \in E(T)$
Now, note that if $q$ is a leaf of $T$ (i.e.~it has no children of its own), then the entire dependence of expression Eq.~(\ref{eq: cpi bound 1}) on $i_q$ is contained in the integral
\begin{equation}
    \int_{-\infty}^{\infty} dE_{i_q} \, e^{S(E_{i_q})- S(E_{i_p i_q})} | f_{i_p i_q} |^2 \frac{\sin(\omega_{i_p i_q} t/2)^2}{\omega_{i_p i_q}^2} \leq \frac{F t}{8},
\end{equation}
which we have already upper bounded in Eq.~(\ref{eq:bound G}).
Applying this upper bound to each leaf of $T$ produces an expression analogous to Eq.~(\ref{eq:bound G}), but where each leaf vertex of $T$ is deleted and replaced a factor of $Ft/8$.
We can then repeat this procedure until the tree is reduced to a single root vertex, at which point the only integral that remains, $\int dE_{i_R} e^{S(E_{i_R})}$, evaluates to the Hilbert space dimension $d$.
This gives
\begin{equation} \label{eq: cpi bound 2}
\begin{split}
    c_\pi 
    & \leq d \, \left( \frac{F t}{8} \right)^m \left( \frac{\mathfrak{m} t}{2} \right)^{2g_\pi},   \\
\end{split}
\end{equation}
which is our final bound on $c_\pi$.

We can now bound the sum over all pairings.
We have
\begin{equation} \label{eq: mu2m bound 1}
    \mu'_{2m} = \sum_\pi c_\pi \leq d \, (F t/8)^m \sum_{g \geq 0} \varepsilon_g(m) \,w^{-2g},
\end{equation}
where $\varepsilon_g(m)$ is the number of pairings of $2m$ elements with genus $g$. Here we define
\begin{equation}
    w = \left( \frac{\mathfrak{m} t}{2} \right)^{-1}.
\end{equation}
The above sum is familiar in random matrix theory and is upper bounded by~\cite{mingo2017free}
\begin{equation}
    \sum_{g \geq 0} \varepsilon_g(m) \,w^{-2g} \leq C_m \exp( m^3 / 2 w^2 ),
\end{equation}
where $C_m$ are the Catalan numbers.
Combining with Eq.~(\ref{eq: mu2m bound 1}), we have our final bound on $\mu'_{2m}$
\begin{equation}
    \mu'_{2m} \leq d \, (F t/8)^m C_m \exp( m^3 / 2 w^2 ) < d \, (F t/2)^m \exp( m^3 / 2 w^2 ),
\end{equation}
where in the latter inequality we use $C_m < 4^m/(m+1)\sqrt{\pi m} < 4^m$~\cite{dutton1986computationally}.
Note that this is equal to the contribution of the leading order Wick contractions [Eq.~(\ref{eq: leading order 2m})] up to an overall factor, $\exp( m^3 / 2 w^2 )$, which is near unity.

We can now show that the probability that $A$ has a maximum eigenvalue that scales as $t$ is exponentially suppressed in the effective Hilbert space dimension.
First, note that the $2m^{\text{th}}$ moment of a matrix $A$ provides an upper bound on the maximum eigenvalue, $\lambda_{\text{max}}^{2m} \leq \mathrm{Trace}( A^{2m} )$.
Now consider the probability that $\lambda_{\text{max}}$ takes value greater than $\sqrt{F t}$ for some constant $\mathfrak{c}$.
We have
\begin{equation}
    P\left[ \lambda_{\text{max}} > \mathfrak{c} \sqrt{F t/2} \right] \leq P\left[ \mathrm{Trace}( A^{2m} ) > (\mathfrak{c}t)^{2m} \right] \leq \frac{ \mathbb{E} \left[ \mathrm{Trace}( A^{2m} ) \right] }{ (\mathfrak{c}t)^{2m} } < d \left( \frac{1}{\mathfrak{c}^2} \right)^m \exp( m^3 / 2 w^2 ),
\end{equation}
where in the second step we use Markov's inequality.
Taking $m = w^{2/3} = \mathcal{O}(\text{exp}(N))$, we find that the probability is doubly exponentially suppressed in $N$ whenever $\mathfrak{c} > 1$,
\begin{equation}
    P\left[ \lambda_{\text{max}} > \mathfrak{c} \sqrt{F t/2} \right] \leq d \left( \frac{1}{\mathfrak{c}^2} \right)^{w^{2/3}} \sqrt{e}.
\end{equation}
Note that the prefactor $d$ grows singly exponentially  in the system size, so the probability is dominated by the doubly exponential suppression. 
Taking $\mathfrak{c} = \sqrt{2}$ gives Lemma~\ref{lemma:OD}. \qed

\subsection{Upper bounding the rank of the diagonal component}\label{app:ETH_diagonal_Aj_rank}

The goal of this section is to establish the following Lemma.
\begin{lemma} \label{lemma:D}
    Consider the setting in Definition~\ref{def:ETH}, and define the  vector of operators $\bs{A}_{\text{d}}(t)$ as in Eq.~(\ref{eq:Ad}).
    Then $\bs{A}_{\text{d}}(t)$ possesses a rank-$R$ approximation with error $\delta$ (Definition~\ref{def: operator rank}), with $R$ defined via
    \begin{equation}
        R(\delta,t) \leq 2 B |\beta_2-\beta_1| \log_2(t \sqrt{N_p} / \delta)^{\gamma+1}.
    \end{equation}
    Here $B$, $\beta_{1}$, $\beta_{2}$, and $\gamma$ are constants defined in Theorem~\ref{thm:multi_parameter_ETH_no_HL}.
\end{lemma}

\emph{Proof of Lemma~\ref{lemma:D}}---We recall that the component operators of $\bs{A}_d$ take the form:
\begin{equation}
    \bs{A}_{\text{d}}(t) = t \, \sum_i \expval{ \bs{P} }_{\beta(E_i)} \dyad{E_i}.
\end{equation}
The central idea of our proof is to exploit the bounded derivatives of $\expval{ P_a }_{\beta}$ with respect to $\beta$ [i.e.~the assumption Eq.~(\ref{eq:bounded_deriv})] to express $\bs{A}_{d}$ as a sum of a low number of vectors.

We do so by Taylor expanding.
Specifically, we first divide the inverse temperature range $[\beta_1,\beta_2]$ into $|\beta_2-\beta_1|/\Delta$ windows of width $\Delta$.
We then form an approximation, $\bs{Q}(\beta; n , \Delta)$, that is equal to the order-$(n-1)$ Taylor expansion of $\expval{\bs{P}}_{\beta}$ within each window
\begin{equation}
\bs{Q}(\beta; n, \Delta) \equiv \expval{\bs{P}}_{\beta_j} + \ldots +  \frac{(\beta-\beta_j)^{n-1}}{(n-1)!} \partial_\beta^{n-1} \, \expval{\bs{P}}_{\beta_j}, \,\,\,\,\,\,\,\,\,\,\, \beta \in [\beta_j,\beta_{j+1})
\end{equation}
where $\beta_j = \beta_1 + j \Delta$ for $j \in \mathbb{N}$.
The approximation carries two free parameters, $\Delta$ and $n$.
According to the remainder theorem, errors in the approximation are suppressed exponentially in $n$:
\begin{equation}
\left| \expval{P_a}(\beta) - Q_a(\beta; n, \Delta) \right| \leq \Delta^{n} \max_\beta \left[ \frac{\partial_\beta^{n} \expval{P_a}}{n!} \right] \leq \left( n^\gamma B \Delta \right)^{n},
\end{equation}
and therefore,
\begin{equation} \label{eq: bound element error}
\bs{w}^T \left( \expval{\bs{P}}(\beta) - \bs{Q}(\beta; n, \Delta) \right) \leq \sqrt{N_p} \left( n^\gamma B \Delta \right)^{n},
\end{equation}
for any normalized vector $\bs{w}^T \bs{w} = 1$, where $N_p$ is the number of unknown parameters.

Putting the above results together, we have the following low-rank approximation for $\bs{A}_{\text{d}}(t)$:
\begin{equation} \label{eq: Ad low rank approximation}
    \bs{A}_{\text{d}}(t) = t \sum_{j=1}^{\frac{|\beta_2-\beta_1|}{\Delta}} \sum_{m=0}^{n-1} \partial_\beta^m \expval{\bs{P}}_{\beta_j} \cdot \left( \sum_{E_i = E(\beta_j)}^{E(\beta_{j+1})} \frac{(\beta(E_i)-\beta_j)^m}{m!} \dyad{E_i} \right) + \bs{E}_{\text{d}},
\end{equation}
where $\bs{E}_{\text{d}}$ is the approximation error (see below).
Note that the left term in the summand is a vector, while the right term is an operator.
The above expression thus matches Definition~\ref{def: operator rank} of a low-rank approximation, with vectors $\{ \bs{b}_\alpha \} \rightarrow \{ \partial_\beta^m \expval{\bs{P}}(\beta_j) \}$ and operators $\{ B_\alpha \}$ equal to the term in parentheses in Eq.~(\ref{eq: Ad low rank approximation}). 
The approximation has rank
\begin{equation}
   \text{dim}\left( \text{span} \{ \partial_\beta^{m} \expval{\bs{P}}(\beta_i) \} \right) \leq \frac{n |\beta_2-\beta_1|}{\Delta}.
\end{equation}
The error in our approximation is
\begin{equation}
    \bs{E}_{\text{d}} = t \sum_i \left( \expval{\bs{P}}_{\beta(E_i)} - \bs{Q}(\beta_i;n,\Delta)  \right) \dyad{E_i}.
\end{equation}
To quantify the error as in Definition~\ref{def: operator rank}, consider any normalized vector $\bs{w}$.
The operator $\bs{w}^T \bs{E}_{\text{d}}$ is diagonal in the energy eigenbasis, with matrix elements bounded in magnitude by $t \sqrt{N_p} (n^\gamma B \Delta)^n$ [see Eq.~(\ref{eq: bound element error})].
We thus have $\lVert \bs{w}^T \bs{A}_d \rVert_s \leq t \sqrt{N_p} (n^\gamma B \Delta)^n$.
The collection of vectors $\{ \partial_\beta^{m} \expval{\bs{P}}_{\beta_i} \}$ therefore forms a rank-$(n|\beta_1-\beta_2|/\Delta)$ approximation of $\bs{A}_{\text{d}}(t)$ with error $\delta = t \sqrt{N_p} (n^\gamma B \Delta)^n$.

The above bound holds for any choice of parameters $n$ and $\Delta$.
To conclude, we make the particular choice $\Delta = (2 n^\gamma B)^{-1}$.
This provides a low-rank approximation of $\bs{A}_{\text{d}}(t)$ comprised of
\begin{equation}
    R(\delta,t) \leq 2 B n^{\gamma+1} |\beta_2-\beta_1|
\end{equation}
vectors, for $\delta = t \sqrt{M} / 2^n$.
Solving for $n = \log_2(M t / \delta)$ and substituting into the above expression, we have our final bound,
\begin{equation}
    R(\delta,t) \leq 2 B |\beta_2-\beta_1| \log_2(t \sqrt{M} / \delta)^{\gamma+1},
\end{equation}
which holds for any $\delta$. \qed

\subsection{Bound on derivatives of expectation values in local Hamiltonians} \label{app: bounded derivative}

In Theorem~\ref{thm:multi_parameter_ETH_no_HL}, we require that the derivatives of thermal expectation values with respect to the inverse temperature are bounded above by constants independent of the system size.
This behavior is widely observed in physical many-body systems and quantum field theories.
In this section, we prove that this bound is obeyed by any local Hamiltonian in $D$-dimensions that exhibits an exponential decay of two-point correlation functions.

We begin by defining the exponential decay of connected two-point correlation functions as in~\cite{kliesch2014locality}:
\begin{definition} \label{2pt decay}
We say that the two-point correlation functions of a set of operators $\mathcal{O}$ decay exponentially if 
\begin{equation}
 \left| \langle A B \rangle_\beta - \langle A \rangle_\beta \langle B \rangle_\beta \right| < a_\beta \lVert A \rVert_s \lVert B \rVert_s \exp( - d(A,B) / \xi_\beta )
\end{equation}
for all $A, B \in \mathcal{O}$.
Here $d(A,B)$ is the Euclidean distance between the support of $A$ and the support of  $B$, and $a_\beta, \xi_\beta$ are constants that depend only on $\beta$ (as in~\cite{kliesch2014locality}).
\end{definition}
\noindent Exponential decay of correlations has been proven for all operators in one-dimensional systems with finite-range interactions at any non-zero temperature~\cite{bluhm2022exponential}, and for operators with bounded spatial support in $d$-dimensional systems with finite-range interactions above a critical temperature~\cite{kliesch2014locality}.

The aim of this section is to establish the following Lemma.
\begin{lemma} \label{dn bound lemma}
Consider a Hamiltonian on a $D$-dimensional cubic lattice with range $r$ and degree $\kappa$.
The exponential decay of two-point correlation functions as in Definition~\ref{2pt decay} implies that the $n^{\text{th}}$ derivative of thermal expectation value is bounded,
\begin{equation} \label{dn bound}
\big| \partial_\beta^n \langle A \rangle_\beta \big| 
\leq c_\beta \lVert A \rVert_s \left( n^{D+2} \left( \frac{2D}{e^{1/2\xi_\beta}-1} \right)^D \frac{\kappa J}{\log(2)} \right)^n, \\
\end{equation}
for any observable $A$ with range $r$. Here we define $c_\beta =  4 a_\beta \mathfrak{b} \, e^{\lceil r/2 \rceil/\xi_\beta}$, with $a_\beta, \xi_\beta, \mathfrak{b}$ as in Definition~\ref{2pt decay}.
Note that the bound is independent of the system size.
\end{lemma}
\noindent In the above, we define the range, $r$, of a Hamiltonian to be the maximum Euclidean distance between two qubits acted on by the same Hamiltonian term.
We also define the degree, $\kappa$, of a Hamiltonian to be the maximum number of terms acting on a single qubit.

Before proceeding to the proof of Lemma~\ref{dn bound lemma}, let us first establish some basic connections between correlation functions and the derivatives of thermal expectation values.
The first derivative of the thermal expectation value gives
\begin{equation}
\partial_\beta \langle A \rangle_\beta = \partial_\beta \left[ \frac{\mathrm{Trace}( A e^{-\beta H} )}{\mathrm{Trace}( e^{-\beta H} )} \right] = - \langle A H \rangle_\beta + \langle A \rangle_\beta \langle H \rangle_\beta,
\end{equation}
which is precisely equal to the connected two-point correlation function in Definition~\ref{2pt decay}.
The second derivative gives
\begin{equation}
\partial_\beta^2 \langle A \rangle_\beta =
\langle A H^2 \rangle_\beta
- 2 \langle A H \rangle_\beta \langle H \rangle_\beta - \langle A \rangle_\beta \langle H^2 \rangle_\beta
+ 2 \langle A \rangle_\beta \langle H \rangle_\beta \langle H \rangle_\beta,
\end{equation}
which is some particular connected three-point correlation function between $A$ and two copies of $H$.
More systematically, for the $n^{\text{th}}$ derivative, we can calculate
\begin{equation}
\partial_\beta ^n \langle A \rangle_\beta = C^{n+1}_\beta(A,H,\ldots,H)
\end{equation}
where we define the \emph{connected $m$-point correlation function} $C^m_\beta(O_1,\ldots,O_m)$,
\begin{equation} \label{n point correlation}
   C^m_\beta(O_1,\ldots,O_m) = -\sum_{P \in \mathcal{P}_m} (-1)^{m+|P|} \cdot (| P |-1)! \cdot \prod_{\sigma=1}^{|P|} \langle \prod_{i \in P_\sigma} O_i \rangle_\beta
\end{equation}
Here the sum is over partitions, $P$, of the first $m$ positive integers, $\{ 1, \ldots, m \}$.
We denote the set of such partitions as $\mathcal{P}_m$.
A partition, $P = \{ P_1, \ldots, P_{|P|} \}$, is specified by $|P|$ disjoint blocks, i.e.~sets $P_\sigma$ for $\sigma = 1, \ldots, |P|$, such that the union of the $P_\sigma$ comprises the full set, i.e.~$\cup_{\sigma=1}^{|P|} P_\sigma = \{ 1 , \ldots, m \}$.
For example, $P = \{ \{ 0, 3 \}, \{ 2,4,5 \}, \{ 1 \} \}$ is a partition of $\{0,1,2,3,4,5\}$ with $|P| = 3$ blocks.
The number of partitions of $m$ elements into $k$ blocks is given by the Stirling numbers of the second kind, denoted $\mathfrak{s}(m,k)$.
For our purposes, the ordering of operators within the expectation values in Eq.~(\ref{n point correlation}) does not matter, since the thermal density matrix $\rho$ and all copies of $H$ commute.

With this formula in hand, we now establish a simple Lemma, which shows that the exponential decay of two-point correlation functions implies exponential decay of $n$-point correlation functions.
\begin{lemma} \label{n+1 bound lemma}
The exponential decay of two-point correlation functions as in Definition~\ref{2pt decay}, implies exponential decay of $n$-point correlation functions
\begin{equation} \label{n+1 bound}
   | C^{n+1}_\beta(O_1,\ldots,O_{n+1}) | \leq
   a_\beta \mathfrak{b} \frac{n!}{\log(2)^n} \Big( \prod_{i =1}^{n+1} \lVert O_i \rVert_s \Big) e^{- d(O_1,\ldots,O_{n+1}) / \xi_\beta}
\end{equation}
for all $O_1,\ldots,O_{n+1} \in \mathcal{O}$.
Here,
\begin{equation} \label{d n+1}
  d(O_1,\ldots,O_n) = \max_i \min_j d(O_i,O_j),
\end{equation}
$\mathfrak{b}$ is constant, and $a_\beta, \xi_\beta$ are constants from Definition~\ref{2pt decay} which depend only on $\beta$.
\end{lemma}
\noindent In local Hamiltonians, we do not expect the above bound to be tight.
In particular, we conjecture that the prefactor of $n!$ can be eliminated, and that the distance in the exponential decay can be strengthened (previous works have speculated that the relevant distance is equal to the length of the minimal Euclidean Steiner tree between $O_1,\ldots,O_{n+1}$~\cite{avdoshkin2019rate}).
Nevertheless, the above bound is sufficient for our purposes, and is convenient since the exponential decay of two-point correlation functions has  been rigorously proven in several cases~\cite{kliesch2014locality,bluhm2022exponential}.
Looking forward, we expect that tighter bounds above would improve the exponent $\gamma$ in Theorem~\ref{thm:multi_parameter_ETH_no_HL}.

We can now proceed to the proof of Lemma~\ref{n+1 bound lemma}.

\emph{Proof of Lemma~\ref{n+1 bound lemma}---} We first re-write the $n+1$-point correlation function as a sum of connected two-point correlation functions
\begin{equation} \label{n+1 point correlation}
   C^{n+1}_\beta(O_1,\ldots,O_{n+1}) = \sum_{P \in \mathcal{P}_n} (-1)^{n+|P|}  (| P |-1)! \, \sum_{\tau=1}^{|P|} \Bigg( \prod_{\sigma \neq \tau} \langle \prod_{i \in P_\sigma} O_i \rangle_\beta \Bigg) \cdot \Bigg( \langle O_{n+1} \! \prod_{i \in P_\tau} O_i \rangle_\beta - \langle O_{n+1} \rangle_\beta \langle \prod_{i \in P_\tau} O_i \rangle_\beta \Bigg)
\end{equation}
Note that the first sum is over partitions of the first $n$ positive integers (not $n+1$), and that the final factor is equal to a connected two-point correlation function between $O_{n+1}$ and $\prod_{i \in P_\tau} O_i$.
Also, while we have written the above expression in terms of $O_{n+1}$ for convenience, the same expression can be applied to any $O_{m}$ since the $n$-point correlation functions are symmetric under index permutations.

To bound the $n$-point correlation function we first isolate the most distant operator, corresponding to $i^* = \text{argmax}_i \min_j d(O_i,O_j)$.
This operator lies at a distance of at least $d(O_1,\ldots,O_{n+1})$ [see Eq.~(\ref{d n+1})]
from all other $O_j$.
Without loss of generality we permute indices such that $i^* = n+ 1$.
We have
\begin{equation} \label{n+1 math}
\begin{split}
    \bigg| C^{n+1}_\beta(O_1,\ldots,O_{n+1}) \bigg| & \leq
   \sum_{P \in \mathcal{P}_n}  (| P |-1)! \, \sum_{\tau=1}^{|P|} \Bigg| \prod_{\sigma \neq \tau} \langle \prod_{i \in P_\sigma} O_i \rangle_\beta \Bigg| \cdot \Bigg| \langle O_{n+1} \! \prod_{i \in P_\tau} O_i \rangle_\beta - \langle O_{n+1} \rangle_\beta \langle \prod_{i \in P_\tau} O_i \rangle_\beta \Bigg| \\
    & \leq
 \sum_{P \in \mathcal{P}_n}  (| P |-1)! \, \sum_{\tau=1}^{|P|} \Big( \prod_{i =1}^{n+1} \lVert O_i \rVert_s \Big) \Big( a_\beta \, e^{- d(O_{n+1},\prod_{i \in P_\tau} O_i) / \xi_\beta} \Big) \\ 
   & \leq
   a_\beta \Big( \prod_{i =1}^{n+1} \lVert O_i \rVert_s \Big) e^{- d(O_1,\ldots,O_{n+1}) / \xi_\beta} \left( \sum_{|P|=1}^n \mathfrak{s}(n,|P|) \cdot | P |! \right). \\ 
\end{split}
\end{equation}
The latter sum is known in mathematics as an ordered Bell number, and satisfies~\cite{bailey1998number}
\begin{equation}
    \sum_{k=1}^n \mathfrak{s}(n,k) \cdot k! = \frac{n!}{2 \log(2)} \left( \frac{1}{\log(2)} \right)^{n} + \varepsilon(n), \,\,\,\,\,\,\,\, \left|\varepsilon(n)\right| <\frac{\pi}{12} \frac{n!}{(2\pi)^n}.
\end{equation}
The error bound can be absorbed into the prefactor of the first term to give the following bound
\begin{equation}
    \sum_{k=1}^n \mathfrak{s}(n,k) \cdot k! \, \leq \, \mathfrak{b} \, n! \left( \frac{1}{\log(2)} \right)^{n},
\end{equation}
with $\mathfrak{b} = \frac{1}{2 \log(2)} + \frac{\log(2)}{24}$.
Inserting the above into Eq.~(\ref{n+1 math}) produces our final bound, Eq.~(\ref{n+1 bound}). \qed

\emph{Proof of Lemma~\ref{dn bound lemma}}---Expanding the Hamiltonian, $H = \sum_i f_a P_a$, we have
\begin{equation}
\partial_\beta^n \langle A \rangle_\beta = \sum_{a_1,\ldots,a_n} \left( \prod_{i=1}^n f_{a_i} \right) C^{n+1}_\beta(A,P_{a_1},\ldots,P_{a_n}),
\end{equation}
and hence
\begin{equation} \label{dn bound first step}
\begin{split}
\big| \partial_\beta^n \langle A \rangle_\beta \big| & \leq J^n \sum_{a_1,\ldots,a_n} \left| C^{n+1}_\beta(A,P_{a_1},\ldots,P_{a_n}) \right| \\
& \leq a_\beta \mathfrak{b} \lVert A \rVert_s n! \left( \frac{J}{\log(2)} \right)^n  \sum_{a_1,\ldots,a_n} e^{- d(A,P_{a_1},\ldots,P_{a_n}) / \xi_\beta} \\
& = a_\beta \mathfrak{b} \lVert A \rVert_s n! \left( \frac{J}{\log(2)} \right)^n  \sum_d \gamma_A(d) e^{- d / \xi_\beta} \\
\end{split}
\end{equation}
where $\gamma_A(d)$ is the number of choices of $\{ a_1, \ldots , a_n \}$ with $d(A,P_{a_1},\ldots,P_{a_n}) = d$.

We define $K(d)$ to be the number of Hamiltonian terms that lie within a distance $d$ of any given Hamiltonian term.
We have the upper bound $K(d) < \kappa (2d + r)^D $.
We can use $K(d)$ to upper bound the distance $\gamma_A(d)$.
Specifically, we have
\begin{equation}
    \sum_{d' = 0}^d \gamma_A(d') \leq n! \, K(d)^n = n! \, \kappa^n \, (2d+r)^{nD}.
\end{equation}
The LHS is equal to the total number of choices of $\{ a_1, \ldots , a_n \}$ with $d(A,P_{a_1},\ldots,P_{a_n}) \leq d$.
To derive the RHS, note that first operator $P_{a_1}$ must lie within distance $d$ of $A$, the second operator must lie within distance $d$ of either $A$ or $P_{a_1}$, and the $m^{\text{th}}$ operator must lie within distance $d$ of at least one of $A$, $P_{a_1}, \ldots P_{a_{m-1}}$.
Since $e^{-d/\xi_\beta}$ is strictly decreasing in $d$, we have
\begin{equation}
\begin{split}
    \sum_d \gamma_A(d) e^{- d / \xi_\beta} & \leq n! \kappa^n \sum_{d=0}^\infty \left[ (2d+r)^{nD} - (2(d-1)+r)^{nD} \right] e^{- d / \xi_\beta} \\ 
    & \leq n! \kappa^n \left( e^{1/\xi_\beta} - 1 \right) \sum_{d=0}^\infty (2d-2+r)^{nD} e^{- d / \xi_\beta} \\ 
    & \leq n! (2^D \kappa)^n e^{\lceil r/2 \rceil/\xi_\beta} \left( 1 - e^{-1/\xi_\beta} \right) \sum_{x=\lceil r/2 \rceil -1}^\infty x^{nD} e^{- x / 2 \xi_\beta} \\ 
\end{split}
\end{equation}
The sum can be expressed in terms of Eulerian numbers and upper bounded,
\begin{equation}
     \sum_{x=0}^\infty x^{nD} e^{- x / 2 \xi_\beta} = \frac{e^{1/2\xi_\beta}}{(e^{1/2\xi_\beta}-1)^{nD+1}} \sum_{i=1}^{nD} A(nD,i) e^{-(i-1)/2\xi_\beta} \leq (nD)! \frac{e^{1/2\xi_\beta}}{(e^{1/2\xi_\beta}-1)^{nD+1}},
\end{equation}
which gives
\begin{equation}
    \sum_d \gamma_A(d) e^{- d / \xi_\beta} \leq 2 \, (nD)! \, (n)! \left( \frac{2^{D} \kappa}{(e^{1/2\xi_\beta}-1)^{D}} \right)^n e^{\lceil r/2 \rceil/\xi_\beta} \frac{1 - e^{-1/\xi_\beta}}{1 - e^{-1/2\xi_\beta}}.
\end{equation}
Inserting the above bound into Eq.~(\ref{dn bound first step}) and using inequalities, $(1 - e^{-1/\xi_\beta})/(1 - e^{-1/2\xi_\beta}) \leq 2$ and $x! \leq x^x$, gives Eq.~(\ref{dn bound}). \qed

\subsection{Learning thermalizing Hamiltonians with $\mathcal{O}(1)$ unknown parameters}\label{app:ETH_single_parameters}

In many Hamiltonian learning scenarios, the number of unknown parameters, $N_p$, scales with some power of the system size, $N$.
In the prior sections, we showed that learning at the Heisenberg limit is not possible in this scenario, if the Hamiltonian thermalizes and one does not have sufficient quantum control.
In this section, we turn instead to learning in thermalizing Hamiltonians with a constant number of unknown parameters, $N_p = \mathcal{O}(1)$.
That is, the Hamiltonians we consider take the form
\begin{equation} \label{eq:H single unknown}
H(u) = u P + \sum_{a=1}^{N_k} v_a Q_a,
\end{equation}
where $u$ is the single unknown parameter, $v_a$ are the known parameters (of total number $N_k$), and $P$, $Q_a$ are $k$-local Pauli operators.
Few-parameter learning problems are potentially easier than many-parameter learning problems, since, in the latter, a lack of knowledge about a given parameter can inhibit learning of other parameters.
We will show that for thermalizing Hamiltonians this is indeed the case, and that a constant number of unknown parameters can, in principle, be learned at the Heisenberg limit even in the absence of quantum control.
Nonetheless, we find that the measurements required to perform such learning may be difficult in practice.

For brevity, we focus on the case of a single unknown parameter, $u$, in the no quantum control model.
Our main results are summarized by the following theorem:
\begin{theorem}
    Consider the learning problem (Definition~\ref{def:learning_problem_formal}) for an $N$-qubit Hamiltonian as in Eq.~(\ref{eq:H single unknown}) in the no quantum control model (Definition~\ref{def:no_quantum_control}).
    Suppose that the Hamiltonian $H(u)$ obeys the eigenstate thermalization hypothesis (Definition~\ref{def:ETH}) within an energy range $[E_1,E_2]$.
    Then:
    \begin{enumerate}
    
    \item Suppose that there exist energies $E_3,E_4 \in [E_1,E_2]$ such that $| \expval{ P }_{\beta(E_3)} - \expval{ P }_{\beta(E_4)} | \geq c$ for a constant $c = \Omega(1)$ independent of the system size. Then there exists a state for which the quantum Fisher information of $u$ scales as $\mathcal{I}^{(Q)} = \Theta(t^2)$.
    
    \item Suppose that the Hamiltonian has bounded degree $\kappa$ and interaction strengths $|u_a|, |v_a| \leq J = \mathcal{O}(1)$, and has bounded derivatives,
    \begin{equation} \label{eq:bounded first energy deriv}
        \left| \partial_E \expval{P}_{\beta(E)} \right| \leq \frac{\tilde{B}}{N}.
    \end{equation}
    Further suppose that the initial state of each experiment is related to a product state by a finite-depth unitary quantum circuit and has energies within $[E_1,E_2]$.
    Then learning the parameter $u$ at the Heisenberg limit incurs an overhead proportional to the square root of the system size, $T = \Omega(\sqrt{N}/\epsilon)$.
    
    \end{enumerate}
\end{theorem}

\emph{Proof---}
We prove both statements 1 and 2 using the quantum Fisher information.
For a single unknown parameter, the quantum Fisher information matrix is simply a positive number.
For pure initial states $\rho_x$ and no quantum control, it takes the form
\begin{equation}
\mathcal{I}^{(Q)}=4\sum_x \left( \mathrm{Trace}[\rho_x A_x^2]-\mathrm{Trace}[\rho_xA_x]^2 \right), \,\,\,\,\,\,\,\,\,\, A_x = A_H(t_x) = \int_0^{t_x} ds\, P(s).
\end{equation}
We refer to Appendix~\ref{app:background} for full details.
As in the multi-parameter case, we decompose $A_H(t) = A_{\text{d}}(t) + A_{\text{od}}(t) + \delta A(t)$ as in Eq.~(\ref{eq: A decomp}).
Again, the latter two terms cannot contribute to the Heisenberg limit owing to Lemma~\ref{lemma:OD} and Definition~\ref{def:ETH}, respectively.
The first term is diagonal in the energy basis, with matrix elements $\bra{E_i} A_d(t) \ket{E_i} = t \expval{P}_{\beta(E_i)}$.
Writing the diagonal matrix elements of $\rho_x$ as $\bra{E_i} \rho \ket{E_i} = p_{x,i}$, we thus have
\begin{equation}
\mathcal{I}^{(Q)}= 4 \sum_x t_x^2 \left( \left[ \sum_i p_{x,i}\expval{P}_{\beta(E_i)}^2 \right] - \left[ \sum_i p_{x,i} \expval{P}_{\beta(E_i)} \right]^2 \right) + \mathcal{O}(t^{3/2}).
\end{equation}
The term in parentheses is the variance of $\expval{P}_{\beta(E_i)}$ over the probability distribution $p_i$.
The sub-leading contribution, $\mathcal{O}(t^{3/2})$, arises from covariances between the diagonal and off-diagonal components of $A_H(t)$.

The quantum Fisher information clearly scales as $t^2$ whenever the above variance is non-zero.
This is possible whenever the thermal expectation value of $P$ varies a non-zero amount with the energy.
This is precisely the condition assumed in statement 1 of the theorem.
From the assumption in statement 1, we can construct a simple state
\begin{equation} \label{eq:rho34}
    \rho = \dyad{\psi_{34}}, \,\,\,\,\,\,\,\,\, \ket{\psi_{34}} = \frac{\ket{E_3}+\ket{E_4}}{\sqrt{2}},
\end{equation}
which features a quantum Fisher information
\begin{equation}
    \mathcal{I}^{(Q)}_x = t_x^2 \left( \expval{P}_{\beta(E_4)} - \expval{P}_{\beta(E_3)} \right)^2 + \mathcal{O}(t_x^{3/2}).
\end{equation}
This proves statement 1. We mention more realistic states that can achieve a quadratic Fisher information in the discussion following this proof.

To prove statement 2, we would like to upper bound the quantum Fisher information in the specific case that $\rho$ is a related to a product state by a finite-depth unitary quantum circuit,
\begin{equation}
    \rho = V \dyad{0} V^\dagger,
\end{equation}
where $\ket{0}$ is the all zeroes state and $V$ is a finite-depth unitary.
We will achieve this by upper bounding the energy variance of the state $\rho$, and using condition Eq.~(\ref{eq:bounded first energy deriv}) to then upper bound the variance of $\expval{P}_{\beta(E_i)}$.

To bound the energy variance, we first consider the rotated Hamiltonian, $H^V = u P^V + \sum_a v_a Q_a^V$, where we denote $O^V = V O V^\dagger$ for an operator $O$.
Note that if $V$ has depth $d$ and $H$ is $k$-local with degree $\kappa$, then $H^V$ is $\tilde{k}$-local with degree $\tilde{\kappa}$, with $\tilde{k} = 2^d k, \tilde{\kappa} = 2^d \kappa$.
Now consider the energy variance of $H$ in the state $V \ket{0}$
\begin{equation}
\delta E^2 \equiv \bra{0} V H^2 V^\dagger \ket{0} - \bra{0} V H V^\dagger \ket{0}^2,
\end{equation}
which is equal to the variance of $H^V$ in the state $\ket{0}$.
We have
\begin{equation}
    \delta E^2 = \sum_{a,b=0}^{N_k} v_a v_b \Big( \bra{0} Q^V_a Q^V_b \ket{0} - \bra{0} Q^V_a \ket{0} \bra{0} Q^V_b \ket{0} \Big),
\end{equation}
abbreviating $u \equiv v_0, P \equiv Q_0$.
The summand above vanishes unless $Q_a$ and $Q_b$ share support.
Each $Q_a$ has support on at most $\tilde{k}$ sites, and thus overlaps with at most $\tilde{k} \tilde{\kappa}$ operators $Q_b$.
We therefore have
\begin{equation}
    \delta E^2 \leq J^2 4^d k \kappa N_k \leq J^2 4^d k \kappa^2 N.
\end{equation}
where in the latter inequality, we use that the bounded degree $\kappa$ upper bounds $N_k \leq \kappa N$.
We thus have our desired result, that the energy variance is $\mathcal{O}(N)$.

We now return to the upper bounding the quantum Fisher information.
Since $\expval{P}_{\beta(E)}$ is a continuous and differentiable function of the energy $E$, we have
\begin{equation}
    \sum_i p_{x,i}\expval{P}_{\beta(E_i)}^2  - \left( \sum_i p_{x,i} \expval{P}_{\beta(E_i)} \right)^2 \leq \left( \partial_E \expval{P}_{\beta(E)} \right)^2 \left[ \sum_i p_{x,i} E_i^2  - \left( \sum_i p_{x,i} E_i \right)^2 \right] = \left( \partial_E \expval{P}_{\beta(E)} \right)^2 \delta E^2.
\end{equation}
Our upper bound on the energy variance, combined with the assumption in statement 2 of the theorem, give the following bound on the quantum Fisher information
\begin{equation}
    \mathcal{I}^{(Q)}_x \leq 4 \sum_x t_x^2 \left( \frac{\tilde{B}}{N} \right)^2 J^2 4^d k \kappa^2 N + \mathcal{O}(t_x^{3/2}) = \mathcal{O}(T^2/N).
\end{equation}
The Cramer-Rao bound now gives our final result, that learning $u$ at the Heisenberg limit requires total evolution time $T = \sum_x t_x = \Omega(\sqrt{N}/\epsilon)$. \qed

In the above proof, we presented a conceptually simple state [Eq.~(\ref{eq:rho34})] that can achieve a quadratically-growing quantum Fisher information.
The state is a superposition of two energy eigenstates of the Hamiltonian, which is unrealistic in practice since eigenstates are hard to prepare.
We now argue that a much more general class of initial states and measurements can achieve a similar Fisher information.
The initial states will be `GHZ-like', in that they are a superposition of states with extensively different energies.
The final measurements will involve backwards time-evolution under a best estimate of the Hamiltonian, as first proposed in Refs.~\cite{wiebe2014hamiltonian,wiebe2015quantum}.

To see this, suppose that the all zero product state $\ket{0\ldots0}$, differs in energy by an extensive amount from the all one product state, $\ket{1\ldots1}$.
This implies that the two states will thermalize to different inverse temperatures, $\beta_{0}$ and $\beta_{1}$, after time-evolution.
Generically, the thermal expectation value of $P$ will differ by an order one amount between $\beta_{0}$ and $\beta_{1}$.
If we take the initial state of the experiment to be a GHZ state,
\begin{equation}
    \rho = \dyad{\text{GHZ}}, \,\,\,\,\,\,\,\,\,\,\,\, \ket{\text{GHZ}} = \frac{\ket{0\ldots0} + \ket{1\ldots1}}{\sqrt{2}},
\end{equation}
this leads to a Fisher information that grows quadratically in time, $\mathcal{I}^{(Q)} \approx t_x^2 ( \expval{P}_{\beta_{0}} - \expval{P}_{\beta_{1}} )^2$.
We note that the temperatures can be adjusted by choosing different initial product states.

It remains to specify what measurement basis is needed to achieve this Fisher information.
As in quantum sensing with GHZ states, the measurement basis will need to be sensitive to the coherence between the all zero and all one state after time-evolution.
That is, we would like to measure the operator
\begin{equation}
    e^{-iH(u)t}\dyad{0 \ldots 0}{1 \ldots 1} e^{iH(u)t} + h.c.
\end{equation}
If the Hamiltonian were known, this could be achieved by first performing backwards time-evolution, i.e. applying $e^{i H(u) t}$, and then measuring the operator $\dyad{0 \ldots 0}{1 \ldots 1}$.

In practice, we expect that backwards time-evolution under the unknown Hamiltonian can be replaced with backwards time-evolution under a best estimate of the Hamiltonian, $H(\tilde{u})$.
If the estimate $\tilde{u}$ has error $\epsilon$, then the backwards time-evolution under $\tilde{u}$ will approximate that under $u$ up to times $t \approx \epsilon$.
This should be sufficient to improve the estimate $\tilde{u}$ by a constant factor.
The experiments can then be updated adaptively as learning proceeds.
This idea was proposed under the moniker `quantum-assisted' Hamiltonian learning in Refs.~\cite{wiebe2014hamiltonian,wiebe2015quantum}, where `quantum-assisted' refers to the ability to backwards time-evolve under $H(\tilde{u})$.
Our analysis thus provides a more thorough justification for the ability of the algorithms in these works to achieve the Heisenberg limit, in the case that the Hamiltonian has a single unknown parameter.
On the other hand, our Theorem~\ref{thm:multi_parameter_ETH_no_HL} establishes that these schemes \emph{cannot} always learn at the Heisenberg limit when the number of unknown parameters is extensive.

\end{document}